\documentstyle[onecolumn,epsfig]{mn}
\def\be{\begin{equation}}
\def\ee{\end{equation}}
\def\simlt{\mathrel{\hbox{\rlap{\hbox{\lower3pt\hbox{$\sim$}}}\hbox{$<$}}}}
\def\simgt{\mathrel{\hbox{\rlap{\hbox{\lower3pt\hbox{$\sim$}}}\hbox{$>$}}}}

\def\age{_{\rm age}}
\def\b{_{\rm b}}
\def\c{_{\rm c}}

\def\coll{_{\rm coll}}
\def\df{_{\rm df}}
\def\drain{^{\rm drain}}
\def\e{_{\rm e}}

\def\E{{\cal E}}
\def\h{_{\rm h}}
\def\H{_{\rm H}}
\def\Hubble{_{\rm Hubble}}
\def\lc{_{\rm lc}}
\def\lr{_{\rm lr}}
\def\lw{_{\rm lw}}
\def\peak{_{\rm peak}}
\def\r{_{\rm r}}
\def\s{_{\rm s}}
\def\t{_{\rm t}}
\def\tri{_{\rm tri}}

\def\d{{\rm d}}
\def\oa#1{\bar #1}
\def\iso#1{\bar #1}
\def\kms{{\rm\,km\,s^{-1}}}
\def\kpc{{\rm\,kpc}}
\def\msun{{\rm\,M_\odot}}
\def\Lsun{{\rm\,L_\odot}}
\def\rsun{{\rm\,R_\odot}}
\def\pc{{\rm\,pc}}
\def\cm{{\rm\,cm}}
\def\yr{{\rm\,yr}}

\def\bfx{{\bf x}}
\def\bfv{{\bf v}}

\def\bfx{{\bf x}}

\title{Evolution of massive binary black holes}
\author[Q. Yu]{Qingjuan Yu\\
Princeton University Observatory, Peyton Hall, Princeton, NJ~08544-1001, USA\\
E-mail: yqj@astro.princeton.edu
}

\begin{document}
\maketitle

\label{firstpage}

\begin{abstract}

\noindent
Since many or most galaxies have central massive black holes (BHs),
mergers of galaxies can form massive binary black holes (BBHs).
In this paper, we study the evolution of massive BBHs in realistic galaxy
models, using a generalization of techniques used to study tidal disruption
rates around massive BHs.
The evolution of BBHs depends on BH mass ratio and host galaxy type.
BBHs with very low mass ratios (say, $\la 0.001$) are hardly ever formed by
mergers of galaxies because the dynamical friction timescale is too long for
the smaller BH to sink into the galactic center within a Hubble time.
BBHs with moderate mass ratios are most likely to form and
survive in spherical or nearly spherical galaxies and in high-luminosity
or high-dispersion galaxies; they are most likely to have merged in
low-dispersion galaxies (line-of-sight velocity dispersion $\la90\kms$)
or in highly flattened or triaxial galaxies.

The semimajor axes and orbital periods of surviving BBHs are generally in the
range $10^{-3}$--$10\pc$ and 10--$10^5\yr$; and they are larger in
high-dispersion galaxies than in low-dispersion galaxies, larger in nearly
spherical galaxies than in highly flattened or triaxial galaxies,
and larger for BBHs with equal masses than for BBHs with unequal masses.
The orbital velocities of surviving BBHs are generally in the range
$10^2$--$10^4\kms$. The methods of detecting surviving BBHs are also discussed.

If no evidence of BBHs is found in AGNs, this may be either because gas plays
a major role in BBH orbital decay or because nuclear activity switches on soon
after a galaxy merger, and ends before the smaller BH has had time to spiral
to the center of the galaxy.

\end{abstract}


\section{Introduction}\label{sec:intro}

\noindent 
It is believed that many or most galaxies house massive black holes (BHs) at
their centers (e.g. Magorrian et al. 1998). Mergers of galaxies, which are a
central
part of the galaxy formation process in the hierarchical structure formation
picture, will inevitably form binary black holes (BBHs) if every galaxy has a
central BH and the BH inspiral time is less than a Hubble time.  The questions
of whether the binaries merge and how long they survive before the merger are
relevant to resolving a number of problems in extragalactic astronomy, such as
the detection of gravitational waves (e.g. Folkner 1998), the bending and
precession of radio jets from active galactic nuclei (AGNs) \cite{BBR80},
the shapes of Fe K$\alpha$ emission line profiles from AGNs \cite{YL01} etc.
On the other hand, so far there is not much observational evidence that BBHs
exist, which might be expected if (i) the lifetime of the BBH is much shorter
than the Hubble time, (ii) the hierarchical model of galaxy formation is
incorrect, or (iii) the BHs are efficiently ejected from the galaxy by
three-body interactions, or other processes.  The purpose of this paper is to
study the lifetime of BBHs in realistic galaxy models.

Consider two massive BHs with masses $m_1\ge m_2$ in the core of a galaxy.
The merger of the BBH mainly involves four processes \cite{BBR80}.
First, the dynamical friction stage: each BH sinks independently towards the
center of the common gravitational potential on the Chandrasekhar dynamical
friction timescale:
\be
t_{\rm df}\sim\frac{4\times10^6}{\log N}\left(\frac{\sigma\c}{200\kms}\right)
\left(\frac{r\c}{100\pc}\right)^2\left(\frac{10^8\msun}{m_i}\right) 
\yr \qquad (i=1,2),
\label{eq:tdf}
\ee 
where the core is assumed to have a one-dimensional velocity dispersion
$\sigma\c$, a core radius $r\c$ and to contain $N$ stars.  Second, the
non-hard binary stage\footnote{This stage is not called the ``soft'' binary
stage because as Quinlan (1996) pointed out, the term ``soft'' is best
restricted to the familiar sense given by Heggie's law, ``soft binaries grow
softer'' (i.e. gain energy and semimajor axis) and a massive BBH in a galaxy
core never grows softer.}: the two BHs eventually form a bound binary system
that continues to lose energy by dynamical friction.
As the binary shrinks, the effectiveness of dynamical friction slowly declines
with the increase of the BH velocity and the shortening of its orbital period,
and the three-body interactions between the BBH and the stars passing in their
vicinity gradually become the dominant factor that makes the BBH lose energy.
Third is the hard binary stage: when 
\be 
a\simeq a\h\equiv\frac{Gm_2}{4\sigma\c^2}=2.8\left({m_2 \over
10^8\msun}\right)\left({200\kms \over \sigma\c}\right)^2\pc
\label{eq:ah}
\ee 
($G$: gravitational constant), the BBH becomes hard \cite{Q96}.  Hard
BBHs lose energy mainly by interacting with stars passing in their vicinity,
most of which will be expelled from the BBH with an energy gain after one or
more encounters with it. 
In such expulsions the average relative change in the BBH energy, $\Delta E/E$,
is independent of $E$ \cite{H75}.
The final stage is the gravitational radiation stage: as the BBH continues to
harden, its semimajor axis decreases to the point at which gravitational
radiation becomes the dominant dissipative force.
A BBH on a circular orbit will then merge within the time \cite{P64}: 
\be
t_{\rm merge}(a)\sim\frac{5}{256}\frac{c^5a^4}{G^3\mu_{12} M_\bullet^2}
=5.8\times10^6\left(\frac{a}{0.01\pc}\right)^4
\left(\frac{10^8\msun}{m_1}\right)^3 \frac{m_1^2}{m_2(m_1+m_2)}\yr,
\label{eq:tmerge}
\ee 
where $\mu_{12}\equiv m_1m_2/(m_1+m_2)$ and $M_\bullet\equiv m_1+m_2$ are the
reduced mass and total mass of the binary, respectively, and $m_2\le m_1$.
If $m_2/m_1\ll1$ (hence, $a\h$ in equation \ref{eq:ah} is small),
gravitational radiation can become the dominant dissipative force before the
hard binary stage begins (see Fig.~\ref{fig:timesph} or
\S~\ref{sec:testassmp} below).

The largest uncertainty in the BBH lifetime comes during the second (non-hard
binary) stage or the third (hard binary) stage, either of which can be the
slowest of the above four processes.
The time spent in the second or third stage may vary by several orders of
magnitude depending on whether or not the low-angular momentum core stars that
interact most strongly with the BBH are depleted before the BBH decays to the
gravitational radiation stage.

In this paper, we will estimate BBH evolution timescales in realistic galaxy
models obtained from observations of the central regions of nearby galaxies. 
To determine the BBH evolution timescale at the slowest stage, the crucial
thing is to get the BBH hardening timescale associated with interactions with
stars in the hard binary stage (see equation \ref{eq:th} below),
which is longer than the hardening timescale in the non-hard binary stage.
The BBH hardening timescale in the hard binary stage gives the slowest
evolution timescale or its upper limit no matter whether the slowest stage
is the hard binary stage or the non-hard binary stage.
So, our analysis starts from the beginning of the third of the above stages,
when the BBH becomes hard.
The theory we shall develop in \S~\ref{sec:lr} and \S~\ref{sec:evlafh} has
been applied to study tidal disruption rates around massive BHs, the evolution
of galactic nuclei, etc. and was first developed to describe the evolution of
globular clusters containing massive black holes \cite{LS77,CK78,S85}.
A simple analysis of the timescales in the first two stages is given in
\S~\ref{sec:evlbfh}.
We will only consider purely stellar dynamical processes, ignoring the
uncertain, but usually small, contribution from gas.
We will also ignore the response of the core structure to the BBH evolution
and assume that the stellar system is in a fixed steady state.
The justification for this assumption is that the total stellar mass removed
from the galactic core during the BBH hardening is generally less than the core
mass (see equations \ref{eq:minact}, \ref{eq:Jrm} or Figures~\ref{fig:mlrjrm},
\ref{fig:minact} below). We present the results of the evolution of BBHs in
realistic galaxy models in \S~\ref{sec:res} and the estimated observational
properties of surviving BBHs in \S~\ref{sec:bbhpro}.
Finally, discussion and conclusions are given in \S~\ref{sec:discon}.

\section{Loss region}\label{sec:lr}

\noindent
In this section, we introduce the concept of ``loss region'' for stellar
systems with BBHs; more specifically, we shall use the terms ``loss cone''
for spherical systems and ``loss wedge'' for axisymmetric systems.

\subsection{Spherical systems}\label{sec:losssph}
\noindent
In a stellar system with a single stellar mass $m_*$ and radius $r_*$,
the distribution function (DF) $f(\bfx,\bfv)$ is defined so that
$f(\bfx,\bfv)\,\d^3\bfx\d^3\bfv$ is the number of stars within a 
phase-space volume $\d^3\bfx\d^3\bfv$ of $(\bfx,\bfv)$.
We will consider the generalization to a distribution of masses and radii
in \S~\ref{sec:gendis}. By Jeans's theorem, in a spherical system the
DF depends on $(\bfx,\bfv)$ only through the orbital binding energy per unit
mass $\E=\psi(\bfx)-{1\over 2}v^2$ and angular momentum per unit mass
$J=|\bfx \times \bfv|$, where $\psi(r)$ is related to the gravitational
potential $\Phi(r)$ through $\psi(r)\equiv-\Phi(r)$.
If a BH has mass $M_\bullet$, stars that pass within a distance
$r_{\rm t}\sim(M_\bullet/m_*)^{1/3}r_*$ of the BH will be tidally
disrupted if $r_{\rm t}\ga r_{\rm s}\equiv 2GM_\bullet/c^2$ 
and swallowed whole if $r_{\rm t}\la r_{\rm s}$, where 
$c$ is the speed of light, and $r_{\rm s}$
is the Schwarzschild radius. 
The region in the (specific energy, specific angular momentum) phase space
where stars can be tidally disrupted at pericenter is called the ``loss cone''
(cf. Frank \& Rees 1976), and given by: 
\be 
J^2\le J\lc^2(\E)\equiv 2r_{\rm t}^2\left[\psi(r_{\rm t})-\E\right] 
\simeq 2GM_\bullet r_{\rm t} \qquad (\E\ll GM_\bullet/r_{\rm t}).
\label{eq:rtlosscone} 
\ee

Now, consider a hard BBH with the masses $m_1\ge m_2\gg m_*$ and semimajor
axis $a\la a\h$ (see equation \ref{eq:ah}).
The radius of the sphere of influence of the BH, $a\H$, is defined implicitly
in terms of the intrinsic one-dimensional velocity dispersion of the galaxy
$\sigma(r)$ through 
\be
\sigma^2(a\H)=GM_\bullet/a\H\equiv\sigma^2\H.
\label{eq:aH}
\ee 
Note that $a\H$ is different from $a\h=(m_2/4M_\bullet)a\H$ which
represents the semimajor axis at which the BBH becomes hard.
Given a total BBH mass $M_\bullet$, $a\h$ depends on the mass of the smaller
BH $m_2$, while $a\H$ does not.
We set the core velocity dispersion $\sigma\c=\sigma\H$ since in the core,
the one-dimensional velocity dispersion $\sigma(r)$ generally varies only
slowly outside $a\H$ [e.g. the ``$\eta$'' models of the density distribution
for $1\le\eta\le 3$ in Dehnen (1993) and Tremaine et al. (1994)].
At large radii ($r\gg a$), the BBH acts as a central point with mass
$M_\bullet=m_1+m_2$ and the stars move in the potential of the stars and the
central mass point $M_\bullet$, with no systematic loss or gain of energy.
The stars interacting most strongly with the BBH have low angular momenta with
pericenters $\la f_a a$ ($f_a$ is a dimensionless factor $\sim 1$).
They may either lose or gain energy after their first encounters with the hard
BBH, but eventually most of the stars will be expelled from the potential of
BBH with an energy gain \cite{Q96}.
As will be seen from equation (\ref{eq:DeltaE}) below, the energy gain of
a star is usually large enough for it to escape from the core when the
BBH is hard, especially when $a\ll a\h$.
By replacing $r_{\rm t}$ with $f_a a$ in equation (\ref{eq:rtlosscone}), the
loss cone in the BBH system is given as: 
\be 
J^2\le J\lc^2(\E,f_a a)\simeq 2GM_\bullet f_a
a =2G(m_1+m_2)f_a a \qquad (\E\ll GM_\bullet/f_a a);
\label{eq:alosscone}
\ee 
thus $J\lc$ will decrease as the BBH hardens. 
The BBH interacts most strongly with the core stars in the loss cone, and
the stars escaping from the core with energy gain are considered as
being removed from the loss cone.
\footnote{ The assumption that stars escaping from the core are removed from
the loss cone is also made by Begelman et al. (1980), Quinlan (1996), and
others. This assumption is plausible since such stars have much longer periods
and are much more susceptible to external torques than core stars. However, in
some circumstances even a small fraction of stars returning to the loss cone
could significantly enhance the BBH decay rate. Therefore this assumption
deserves further investigation.}
The BBH energy loss rate is determined by the rate of removal of stars from
the loss cone (the ``clearing rate'').
At first, the depletion of the initial population of stars in the
loss cone determines the clearing rate.  As stars are removed from the loss
cone, new stars are scattered into the loss cone by two-body relaxation
\cite{BT87} and resonant relaxation \cite{RT96,RI98},
and eventually the clearing rate reaches a steady state controlled by the
balance between the loss rate and the rate at which stars
refill the loss cone.  If the rms angular momentum transferred to or from the
stars per orbital period is larger than $J\lc$, then the stars will refill the
loss cone as fast as it is depleted and the loss cone remains full; otherwise,
the stars will slowly diffuse into the loss cone and the loss cone remains
nearly empty.  Thus at large radii, the loss cone is full (``pinhole'' regime)
and at small radii, the loss cone is nearly empty (``diffusion'' regime)
[the terms come from Lightman \& Shapiro (1977)].
The transition radius or binding energy between the two regimes is denoted by
$r\lc$ or $\E\lc$ [$\psi(r\lc)\equiv\E\lc$].
If most of the stars contributing to the BBH orbital decay have energy
$\E\ll GM_\bullet/f_a a$, which is generally
a good approximation as will be seen from \S~\ref{sec:testassmp}, the rate of
refilling the loss cone in a BBH system can be determined from existing tidal
disruption calculations, simply by replacing the tidal disruption radius $r_t$
with $f_a a$.  The refilling rate caused by two-body relaxation can be
obtained by solving the steady-state Fokker-Planck equation (e.g. equation 17
in Magorrian \& Tremaine 1999, hereafter MT).  The loss cone in the BBH system
can be approximated as static even though the loss cone shrinks slowly as the
BBH orbit decays because the timescale of the stellar diffusion into the loss
cone is generally shorter than the BBH hardening timescale
(see \S~\ref{sec:testassmp} below).
Resonant relaxation is effective only within a radius enclosing a mass $\sim
0.1M_\bullet$ of stars and does not contribute significantly to the total
refilling rate; therefore, it will be neglected
(see MT or \S~\ref{sec:testassmp} below).

\subsection{Axisymmetric systems and triaxial systems}\label{sec:lossaxis}

\noindent
In an axisymmetric galaxy, the DF involves three integrals: $f(\E,J_z,J_3)$
where $J_z$ is the component of angular momentum along the symmetry ($z$-)axis
and $J_3$ is a third integral.  If the axisymmetric galaxy is nearly
spherical, the third integral $J_3$ may be approximated by the angular
momentum $J$ \cite{BT87}.  In axisymmetric galaxies, there exist
centrophilic orbits such as box orbits, which pass arbitrarily close to
the center and have low angular momentum, as well as centrophobic orbits
such as loop orbits, which avoid the center and have high angular momentum.
Here, we introduce $J\s$ to mark the transition from centrophilic ($J\la J\s$)
to centrophobic ($J\ga J\s$) orbits.
Stars on centrophilic orbits with $|J_z|<J\lc$ can precess into the loss cone,
while those on centrophobic orbits with $J>J\lc$ cannot.
Hence, in an axisymmetric galaxy, the loss cone is replaced by a loss wedge
with $|J_z|<J\lc$ (MT).
The BBH energy loss rate is determined initially by the clearing rate from the
loss wedge and in the steady state by the rates at which two-body
relaxation and tidal forces refill the loss wedge.

Likewise, in a triaxial galaxy, we may expect that there also exists some
characteristic angular momentum $J\s$ which marks the transition from
centrophilic to centrophobic orbits, and most of the stars with $J<J\s$ can
precess into the loss cone. 
We shall call the region $J<J\s$ the ``loss region'' in triaxial galaxies. 
Thus, the BBH energy loss rate is determined initially by the clearing rate
from the loss region and in the steady state by the rates at
which two-body relaxation and tidal forces refill the loss region.

\section{Evolution timescales after BBHs become hard}\label{sec:evlafh}

\noindent
The stars contributing to the orbital decay of a hard BBH come from the loss
region.  Assuming that $F(\E,a)\d\E$ is the clearing rate from the loss region
for stars with energy $\E\rightarrow\E+\d\E$, the energy loss rate of the BBH
is given by: 
\be 
\left|\frac{\d E}{\d t}(a)\right|=-m_*\int\Delta\E F(\E,a)\,\d\E,
\label{eq:energyloss}
\ee
where $\Delta\E$ is the average specific energy change of the stars escaping
from the BBH during an interaction with the BBH, and the semimajor axis
of the BBH is $a$.
The quantity $\Delta \E$ is only weakly dependent on $\E$ for hard binaries.
Hence we assume 
\be
\Delta\E=-KG\mu_{12}/a,
\label{eq:DE}
\ee
where $\mu_{12}$ is the reduced mass of the BBH and $K$ is a constant
which will be determined from Quinlan (1996) in \S~\ref{sec:K}.
Thus, after the BBH becomes hard, its hardening time due to 
stellar interactions is given as:
\be
t\h(a)=\left|\frac{a}{\dot a}\right|=\left|\frac{E}{\dot
E}\right|=\frac{M_\bullet}{2Km_*}\frac{1}{\int F(\E,a)\,\d\E}. 
\label{eq:th}
\ee
The timescale associated with gravitational radiation is given by \cite{P64}:
\be
t_{\rm gr}(a)=\left|\frac{a}{\dot a}\right|=\frac{5}{64}
\frac{c^5a^4(1-e^2)^{7/2}}{G^3\mu_{12}M^2_\bullet(1+73/24e^2+37/96e^4)},
\label{eq:tgr}
\ee
where $e$ is the BBH orbital eccentricity.
The merger timescale in equation (\ref{eq:tmerge}) is related to the timescale
in equation (\ref{eq:tgr}) by
$t_{\rm merge}(a)=\int_0^{a}t_{\rm gr}(a,e=0)\,\d a/a$.

As will be seen in \S~\ref{sec:K}, the hardening timescale $t\h(a)$ of a
hard BBH is independent of its orbital eccentricity $e$.
The gravitational radiation timescale in equation (\ref{eq:tgr}) depends
weakly on $e$ when $e$ is small
[e.g. $t_{\rm gr}(a,e=0.3)/t_{\rm gr}(a,e=0)\simeq 0.6$];
but for large $e$, it significantly decreases
[e.g. $t_{\rm gr}(a,e=0.8)/t_{\rm gr}(a,e=0)\simeq 0.01$].
In this paper, we assume that the initial eccentricity when the BBH becomes
hard is small (say, $e\la 0.3$).
In this case, in the hard binary stage, the BBH eccentricity hardly grows as
the BBH hardens \cite{Q96};
in the gravitational radiation stage, the eccentricity decays
exponentially. So in our calculations we will always set $e=0$.

Combining equations (\ref{eq:th}) and (\ref{eq:tgr}), we find that the
evolution timescale of the hard BBH $t_{\rm evol}(a)$ is given by:
\be
\frac{1}{t_{\rm evol}(a)}=\frac{1}{t\h(a)}+\frac{1}{t_{\rm gr}(a)}
\label{eq:tevol}
\ee
and gravitational radiation becomes dominant when $a\le a_{\rm gr}$, where
$a_{\rm gr}$ is defined by $t\h(a_{\rm gr})=t_{\rm gr}(a_{\rm gr})$.
This marks the transition between the hard binary and gravitational radiation
stage.
The total mass of the stars interacting with the BBH during the hard binary
stage ($a_{\rm gr}\le a\le a\h$), which can be derived from equation
(\ref{eq:th}), is given by:
\be
M_*^{\rm interact}=m_*\int\!\!\!\int F(\E,a)\,\d\E\d t=\frac{M_\bullet}{2K}\ln(\frac{a\h}{a_{\rm gr}}). 
\label{eq:minact}
\ee

\subsection{$K$}\label{sec:K}

\noindent
Quinlan (1996) studied the dynamical evolution of massive BBHs
by scattering experiments using the restricted three-body approximation.
Consider a hard BBH in a galaxy core with uniform density $\rho$ and
Maxwellian velocity dispersion $\sigma\c$. The stars in the core
start at a radius $r\gg a$ from the BBH, with velocity ${\bf v}$. 
Assume that $v\r$ and $v\t$ are the radial and tangential components
of the vector ${\bf v}$. At large radii, the low angular-momentum stars
that contribute to the orbital decay of the hard BBH are in nearly radial
motion and the number flux into the loss cone is:
\be 
\int F(\E,a)\d\E=\int_{0}^\infty f(v\r,v\t=0,r)4\pi^2
J\lc^2(f_a a)|v_{\rm r}|\,\d v_{\rm r},
\label{eq:dEdtQuinlan}
\ee
where
\be
f(v\r,v\t=0,r)=\frac{\rho}{m_*}\frac{1}{(2\pi\sigma\c^2)^{3/2}}
\exp(-v\r^2/2\sigma\c^2). 
\ee
With $J\lc^2(f_a a)\simeq 2GM_\bullet f_a a$ at large radii, we have from
equation (\ref{eq:th}) the hardening timescale of the BBH: 
\be 
t\h(a)=\left|\frac{a}{\dot a}\right|=\frac{1}{4\sqrt{2\pi}Kf_a}\frac{\sigma\c}{Ga\rho}.  
\label{eq:thK} 
\ee 
In Quinlan's simulation, the hardening time is given as
\be 
t\h=\left|\frac{a}{\dot a}\right|=\frac{\sigma\c}{G\rho aH}
=2.8\times10^7\yr\left(\frac{\sigma\c}{200\kms}\right)
\left(\frac{10^3\msun/\pc^3}{\rho}\right)
\left(\frac{0.1\pc}{a}\right)\left(\frac{16}{H}\right),
\label{eq:thQuinlan}
\ee
where for a hard BBH, $H$ is a constant ($\sim 16$) independent of the BH
mass ratio and orbital eccentricity.  Combining equations (\ref{eq:thK}) and
(\ref{eq:thQuinlan}), we have the constant $Kf_a=0.097H=1.56$. In our
calculations, we will always set $K=1.56$ and $f_a=1$.
Thus, combining equations (\ref{eq:ah}) and (\ref{eq:DE}), the average
specific energy change is given by
\be
\Delta\E=K{G\mu_{12}\over a}\simeq 2\left({K\over1.56}\right)\left({2m_1\over m_1+m_2}\right)\left({a\h \over a}\right)\left({3\sigma\c^2\over2}\right),
\label{eq:DeltaE}
\ee
which is generally large enough for the star to escape from the core when
the BBH becomes hard, especially when $a\ll a\h$.

\subsection{$F(\E,a)$}\label{sec:F}

\noindent
Most of this subsection is a summary and generalization of the tidal disruption
rates calculated in MT.

\subsubsection{Spherical galaxies}\label{sec:Fsph}

\noindent
In a spherical galaxy with a DF $f(\E,J^2)$, when the BBH first becomes hard 
(the time $t$ is set to 0), the total stellar mass in the
loss cone is given by:
\be
M\lc(a\h)\simeq m_*\int 4\pi^2\eta(\E)f(\E,J^2=0)J^2\lc(\E,f_a a\h)P(\E)\,\d\E \qquad 0\le \eta(\E)\le 1,
\label{eq:Mlc}
\ee
where $P(\E)$ is the radial period of an orbit with energy $\E$ and
zero angular momentum and $\eta(\E)$ is a dimensionless factor. 
If $\eta(\E)=1$ (or 0) for all $\E$, the loss cone is full (or empty).
The clearing rate per unit energy at time $t$ is first mainly controlled by the
draining rate from the loss cone:
\be
F\drain(\E,a;t)\,\d\E\simeq \cases{ 4\pi^2\eta(\E)
f(\E,J^2=0)J\lc^2(\E,f_a a)\,\d\E & for $t<P(\E)$ \cr
        0		           & for $t>P(\E)$}.
\label{eq:Fdrainsph}
\ee
After time $P(\E)$, when the loss cone of stars with energy $\E$ is
depleted, the loss cone is refilled by two-body relaxation.
The steady-state diffusion rate of stars into the loss cone
is given by the steady-state solution of the Fokker-Planck
equation (13) in MT, which is a generalization of the
equation in Cohn \& Kulsrud (1978) to a non-Keplerian potential:
\be
F^{\rm lc}(\E,a)\,\d\E=\frac{F^{\rm max}(\E)\,\d\E}{\ln R^{-1}_0(\E,f_a a)},
\label{eq:Flc}
\ee
where
\be
F^{\rm max}(\E)\equiv 4\pi^2\oa f(\E)J^2\c(\E)\oa\mu(\E)P(\E),
\ee
\be
R_0(\E,f_a a)\equiv\frac{J\lc^2(\E,f_a a)}{J^2\c(\E)}\times\cases{
      \exp[-q(\E,f_a a)] & for $q(\E,f_a a)>1$\cr
      \exp[-0.186q(\E,f_a a)-0.824\sqrt{q(\E,f_a a)}] & for $q(\E,f_a a)<1$},
\label{eq:R0}
\ee
\be
q(\E,f_a a)\equiv \oa\mu(\E)P(\E)J^2\c(\E)/J\lc^2(\E,f_a a),
\label{eq:qea}
\ee
\be
\oa\mu(\E)\equiv 2\int^{r_+}_{r_-}\frac{\mu\,\d r}{v_{\rm r}}\Big/ P(\E),\qquad
\mu\equiv\frac{2r^2\langle\Delta v_{\rm t}^2\rangle}{J\c^2(\E)},
\label{eq:mu}
\ee
$J\c(\E)$ is the specific angular momentum of a circular orbit
at energy $\E$, $\iso f(\E)$ is the ``isotropized'' DF defined by
$\iso f(\E)\equiv\int_0^{J\c^2}f(\E,J^2)\,\d J^2/J\c^2(\E)$,
$r_+$ and $r_-$ are the apocenter and pericenter of loss-cone orbits,
$\langle\Delta v_{\rm t}^2\rangle$ is the diffusion coefficient for tangential
velocity (see equation 8-64 of Binney \& Tremaine 1987),
and $\oa\mu(\E)$ is the orbit-averaged diffusion coefficient.
Note that the stellar encounters responsible for the velocity diffusion
can be treated as elastic encounters only outside the ``collision radius'':
\be
r\coll\simeq 7\times 10^{10}(M_\bullet/m_*)(r_*/\rsun)\cm
\label{eq:rcoll}
\ee 
at which the velocity dispersion $\sim\sqrt{GM_\bullet/r}$ is comparable
with the escape velocity from typical stars \cite{FR76}.  When $r\la
r\coll$, two stars cannot deflect each other's velocities through a large
angle without coming so close that they actually collide.
Generally, most stars have $R_0(\E,f_a a)\ll 1$, and the isotropized DF
is close to the real stellar DF and is related to the stellar density
$\rho(r)$ by 
\be 
\rho(r)\approx 4\pi m_*\int\iso f(\E)\sqrt{2[\Psi(r)-\E]}\,\d\E.
\label{eq:nuf}
\ee
The transition between pinhole ($\E<\E\lc$) and diffusion ($\E>\E\lc$) regimes
is at $q(\E\lc,f_a a)\simeq -\ln [J\lc^2(\E\lc,f_a a)/J\c^2(\E)]$.
From equations (\ref{eq:Flc})--(\ref{eq:mu}) it can be seen that in the
diffusion regime the flux into the loss cone $F^{\rm lc}(\E,a)$ is insensitive
to $a$ since it depends only logarithmically on $a$.

Considering both the draining and the refilling of the loss cone,
at a certain time $t$, the clearing rate is given as:
\be
F(\E,a;t)=\max [F\drain(\E,a;t),F^{\rm lc}(\E,a)].
\label{eq:Fsph}
\ee

\subsubsection{Axisymmetric galaxies}\label{sec:Faxis}

\noindent
In spherical systems, stars are all on centrophobic loop orbits.
In axisymmetric galaxies, there also exist centrophilic orbits which can
enhance the clearing rates.
The dynamics of eccentric orbits in non-spherical potentials can be studied
by a simple symplectic map constructed by Touma \& Tremaine (1997).
The mapping models the evolution of eccentric orbits as a two-step process:
(i) precession of the orientation of the orbit in a non-spherical potential,
and (ii) a kick to the angular momentum of the orbits at apocenter where
the star spends most of its time, and the torques are likely to be strongest.
Following MT, we use this mapping to study the dynamics of centrophilic
orbits with $J_z=0$ in axisymmetric galaxies:
\begin{eqnarray}
Y'_n & = & Y_n-{1\over2}\epsilon\sin 2\theta_n \nonumber \\
\theta_{n+1} & = & \theta_n + g(Y'_n) \nonumber \\
Y_{n+1} & = & Y'_n -{1\over2}\epsilon\sin2\theta_{n+1}
\label{eq:TTmap}
\end{eqnarray}
where $Y=J_{\pm}/J\c$, $J_{\pm}$ is the scalar angular momentum ($|J_{\pm}|=J$)
which can be positive or negative,
$\theta$ is the colatitude at each apocenter passage,
$\epsilon J\c$ is the time-integral of the torque over one radial period
and $g(Y)$ is the advance in $\theta$ per radial period.
The parameter $\epsilon$ is related to the axis ratio of the potential $b$
and the mass ellipticity $\epsilon'$, for example,
\be
\epsilon=\sqrt{2\pi {\rm e}}(1-b)=\sqrt{2\pi {\rm e}}\epsilon'/3 \qquad (b<1)
\label{eq:epsilonb}
\ee
in the logarithmic potential \cite{TT97}
\be
\Phi={1 \over 2}\log (x_1^2+x_2^2/b^2).
\label{eq:potscfr}
\ee 
Of course, the potential of realistic galaxies with central BHs is not
scale-free, and so the precession rate and the mapping in equation
(\ref{eq:TTmap}) are dependent on the energy $\E$.
As shown in Figure~4 of MT, in realistic galaxies, the centrophilic orbits are
regular at large $\E$ and stochastic below some critical energy
$\E_{\rm crit}$.
Above $\E_{\rm crit}$, the peak angular momenta of the regular orbits passing
the points ($Y=0,\theta=0,\pi$) in the $Y$--$\theta$ surfaces of section are
denoted by $J_{\rm m}(\E)$ and the area of ($Y,\theta$) phase space covered by
the regular orbits with peak angular momenta less than $J_{\rm m}$ is
$4J_{\rm m}(\E)/J\c(\E)$.
Below $\E_{\rm crit}$, the areas of ($Y,\theta$) phase space occupied by the
stochastic orbits are denoted by $2\pi J_l(\E)/J\c(\E)$.
Those stars on centrophilic
orbits with $|J_\pm|<J_{\rm m}$ (for regular orbits) or $|J_\pm|<J_l$ (for
stochastic orbits) can precess into the loss cone so long as $|J_z|<J\lc$.
Here, we use one symbol $J\s(\E)$ to represent $2J_{\rm m}(\E)/\pi$
for regular orbits or $J_l(\E)$ for stochastic orbits.

If the DF of the axisymmetric galaxy is a function of two integrals
$f(\E,J_z)$, when the BBH first becomes hard,
the total mass in the loss wedge is given by:
\begin{eqnarray}
M\lw(a\h)\simeq m_*\int 4\pi^2\eta(\E)f(\E,J_z=0)\max[J^2\lc(\E,f_a a\h),2J\lc(\E,f_a a\h)J\s(\E)]P(\E)\,\d\E \qquad
0\le \eta(\E)\le 1,
\label{eq:Mlw}
\end{eqnarray}
where as in equation (\ref{eq:Mlc}), the dimensionless factor $\eta(\E)=1$
(or 0) for a full (or empty) loss wedge.
If $J\lc(\E,f_a a)$ is larger than $2J\s(\E)$, we use the clearing rates
obtained in the spherical case.
After $J\lc(\E,f_a a)$ becomes less than $2J\s(\E)$ at a time given by
$t\equiv T\s(\E)$ ($t=0$ at $a=a\h$), we will consider the effects of
flattening as follows.  First, the loss wedge in
a two-integral model drains at a rate: 
\be
F\drain(\E,a;t)\,\d\E\simeq\cases{4\pi^2\eta(\E)f(\E,J_z=0)J\lc^2(f_a a)\,\d\E & for $t<T\drain(\E)$ \cr 0 & for $t>T\drain(\E)$},
\label{eq:Fdrainaxis}
\ee
where $T\drain(\E)$ satisfies 
\begin{eqnarray}
\int^{\min[P(\E),T\s(\E)]}_0\d t'(a')~
\frac{2J\lc(f_a a)J\s(\E)}{J^2\lc(f_a a')}F\drain(\E,a';t')\,\d\E
+\int_{\min[P(\E),T\s(\E)]}^{T\drain(\E)}\d t'(a')~
\frac{J\lc(f_a a)J\s(\E)}{J\lc(f_a a')J\s(\E)}F\drain(\E,a';t')\,\d\E
\nonumber \\
\simeq 8\pi^2\eta(\E)f(\E,J_z=0)J\lc(f_a a)J_{\rm s}(\E)P(\E)\,\d\E.
\label{eq:Tdrainlw}
\end{eqnarray}
In the above equation, the right-hand-side term represents the total number
of stars (with energy $\E\rightarrow\E+\d\E$) which can precess into the loss
cone $J<J\lc(f_a a)$ by tidal forces.
If $T\s(\E)<P(\E)$, the left-hand-side terms in equation (\ref{eq:Tdrainlw})
approximately represent the numbers of stars (with energy
$\E\rightarrow\E+\d\E$) removed from the loss wedge before and after
$J\lc(\E,f_a a)=2J\s(\E)$;
and if $T\s(\E)>P(\E)$, we have $T\drain(\E)\simeq P(\E)$.

After the loss wedge is depleted, the rate at which the loss wedge is refilled
is given by a Fokker-Planck analysis in MT:
\be
F^{\rm lw}(\E,a)\d\E=4\pi^2J\lc^2(\E,f_a a){\iso f(\E)\,\d\E
\over 1+J\c/(4q_zJ\s)}, 
\label{eq:Flw}
\ee 
where $q_z=q/8$ (see $q$ in equation \ref{eq:qea})
and $\iso f(\E)$ is the ``isotropized''
DF defined by $\oa f(\E)\equiv\int_0^{J\c}f(\E,J_z)\,\d J_z/J\c(\E)$.
If the galaxy is close to spherical, the isotropized DF $\iso f(\E)$ is close
to the real stellar DF and $\iso f(\E)$ is also
related to the stellar density by equation (\ref{eq:nuf}).
The transition between the ``diffusion'' and ``pinhole'' regions occurs at
$q\simeq 2J\c/J\s$.
From equations (\ref{eq:qea}) and (\ref{eq:Flw}) it can be seen that
in the diffusion regime the flux into the loss wedge $F^{\rm lw}(\E,a)$
is independent of $a$.

Considering both the draining and refilling of the loss wedge or the loss cone,
at a certain time $t$, the clearing rate is given as:
\be
F(\E,a;t)=\max [F\drain(\E,a;t),F^{\rm lw}(\E,a),F^{\rm lc}(\E,a)].
\label{eq:Faxis}
\ee

\subsubsection{Triaxial galaxies}\label{sec:Ftri}

\noindent
Similarly, in triaxial galaxies, there will be some characteristic
angular momentum $J_{\rm s}(\E)$ inside which most orbits are centrophilic and
the BBH decay rate is determined by the clearing rates from the
loss regions $J(\E)<J_{\rm s}(\E)$.
Here, we approximate the $J\s(\E)$ obtained in the axisymmetric cases as the
characteristic angular momentum $J_{\rm s}(\E)$ in triaxial galaxies.
Merritt and Quinlan (1998) argue that the triaxiality of galaxies with central
BHs decays due to the evolution of stochastic orbits, so that most galaxies
become axisymmetric at any given radius within a time
$t\equiv T^{\rm trans}\sim 10^2$ local orbital periods.
When the BBH first becomes hard, we set the time $t=0$ and the age of the
galactic triaxiality is defined as $T^{\rm tri}\age$ at this time.
We give the mass in the loss region at $a=a\h$ as:
\be
M\lr(a\h)\simeq m_*\int 4\pi^2\eta(\E)f(\E)
\max[J\lc^2(\E,f_a a\h),J\s^2(\E)]P(\E)\,\d\E \qquad 0\le \eta(\E)\le 1.
\label{eq:Mlr}
\ee

When $0\le t\le T\tri(\E)\equiv\max[T^{\rm trans}-T^{\rm tri}\age,0]$,
we consider the clearing rates in triaxial galaxies as follows.
If $J_{\rm s}(\E)<J\lc(\E)$, the clearing rates
follow the same formula as in the spherical case (see \S~\ref{sec:Fsph}).
When $J_{\rm s}(\E)$ becomes larger than $J\lc(\E)$ at a time given by
$t\equiv T\s(\E)$, the clearing rates from the 
loss region are first controlled by the draining rates from the loss region:
\be
F\drain(\E,a;t)\,\d\E\simeq\cases{4\pi^2\eta(\E)f(\E)J\lc^2(f_a a)\,\d\E & for $t<\min[T\drain(\E),T\tri(\E)]$, \cr
0 & for $t>\min[T\drain(\E),T\tri(\E)]$}
\label{eq:Fdraintri}
\ee
where $T\drain$ satisfies
\begin{eqnarray}
\int^{\min[P(\E),T\s(\E)]}_0\d t'(a')~
\frac{J^2\s(f_a a)}{J^2\lc(f_a a')}F\drain(\E,a';t')\,\d\E
+\int_{\min[P(\E),T\s(\E)]}^{T\drain}\d t'(a')~F\drain(\E,a';t')\,\d\E
\nonumber \\
\simeq 4\pi^2\eta(\E)f(\E)J^2\s(\E)P(\E)\,\d\E.
\label{eq:Tdrainlr}
\end{eqnarray}
Equation (\ref{eq:Tdrainlr}) has similar meanings to equation
(\ref{eq:Tdrainlw}).

After the loss region is depleted, the refilling of the loss region can be
described by equation (\ref{eq:Flc}) or (\ref{eq:Flw}).
The clearing rate is given as equation (\ref{eq:Faxis}).

When $t\ge T\tri(\E)$, we will assume axisymmetry and follow the
calculation in \S~\ref{sec:Faxis}.

\section{Evolution timescales before BBHs become hard}\label{sec:evlbfh}

\noindent
In this section, we carry out a simple analysis of the evolution timescales
before the BBH becomes hard.  Assume that a BH of mass $m_1$ is located at the
center of a spherical galaxy with stellar mass density $\rho(r)$ and
one-dimensional velocity dispersion $\sigma(r)$.  Suppose that a BH $m_2$
($m_2\le m_1$) that used to be at the center of a galaxy is orbiting with a
velocity $\bfv_2$ and spiraling into the center of the parent galaxy by
dynamical friction.  Because the inspiraling BH $m_2$ is accompanied by the
stars bound to it with total mass $M^{m_2}_*$, dynamical friction brings the
two BHs together much more rapidly than if $m_2$ is ``naked''
(e.g. Milosavljevi\'c \& Merritt 2001).  The dynamical friction force on $m_2$ and its
accompanying stars $M^{m_2}_*$ is given by \cite{BT87}: 
\be
(m_2+M_*^{m_2})\left(\frac{\d \bfv_2}{\d t}\right)\df=-4\pi\ln\Lambda
G^2(m_2+M^{m_2}_*)^2\rho(r)\frac{\bfv_2}{v_2^3} \left[{\rm
erf}(X)-\frac{2X}{\sqrt{\pi}}{\rm e}^{-X^2}\right],
\label{eq:fricforce}
\ee
where $X=v_2/\sqrt{2}\sigma(r)$, erf is the error function, and $m_2$ and its
bound stars $M_*^{m_2}$ are assumed to act as a single body.
The logarithm of the ratio of the maximum and the minimum impact parameter
$\ln\Lambda$ is set to unity (which overestimates the dynamical friction
timescale, but not by more than a small logarithmic factor).
Assuming that the orbit of $m_2$ is circular, the force (equation
\ref{eq:fricforce}) is tangential and causes $m_2$ to lose angular momentum
per unit mass $L$ at a rate:
\be
\frac{\d L(r)}{\d t}=r\left(\frac{\d v_2}{\d t}\right)\df,
\label{eq:dLdt}
\ee
where
\be
L(r)=rv_2=\sqrt{G[m_1+M_*(r)]r}
\label{eq:L}
\ee
and $M_*(r)$ is the stellar mass in the parent galaxy within a radius $r$.
The dynamical friction timescale is given by:
\be
t\df=\left|\frac{r}{\dot r}\right|=r\left|\frac{\d L/\d r}{\d L/\d t}\right|,
\label{eq:frictime}
\ee
which can be calculated from equations (\ref{eq:fricforce})--(\ref{eq:L}).

In the potential of the parent galaxy with BH $m_1$, the stars around the BH
$m_2$ are tidally truncated at the radius 
\begin{eqnarray}
r_t & \simeq &
\left[\frac{G(m_2+M_*^{m_2})}{4\Omega^2-\kappa^2}\right]^{1/3} \nonumber \\ 
& \simeq & \left(\frac{GM_*^{m_2}}{4\Omega^2-\kappa^2}\right)^{1/3}
\qquad M_*^{m_2}\gg m_2,
\label{eq:tidrad}
\end{eqnarray}
where $\Omega$ and $\kappa$ are the circular frequency and epicycle frequency
in the parent galaxy.
Note that the tidal forces approach zero in a nearly homogeneous core,
in which $\kappa=2\Omega$.
We use this formula to obtain a crude estimate of the ratio of the bound
stellar mass $M_*^{m_2}$ to the BH mass $m_2$.
If the distribution of the stars around each of the BHs is a singular
isothermal sphere and their one-dimensional velocity dispersions are
$\sigma_1$ and $\sigma_2$, we have $\Omega^2=2\sigma_1^2/r^2$,
$\kappa^2=4\sigma_1^2/r^2$, $M_*^{m_2}=2\sigma_2^2r_t/G$ and
$r_t=(\sigma_2/\sqrt{2}\sigma_1)r$.
If the velocity dispersions $\sigma_1$ and $\sigma_2$ follow the same relation
with central BH masses $m_1$ and $m_2$ as equation (\ref{eq:msigma})
(see \S~\ref{sec:galpro} below), the ratio of the bound stellar mass
$M_*^{m_2}$ to the BH mass $m_2$ is given by:
\be
\frac{M^{m_2}_*}{m_2}=\frac{\sqrt{2}r\sigma_2^3}{Gm_2\sigma_1}
=300\left(\frac{r}{2\kpc}\right)\left(\frac{m_2}{m_1}\right)^{0.25}
\left(\frac{10^8\msun}{m_1}\right)^{2.14} \qquad M_*^{m_2}\gg m_2.
\label{eq:mboundratio}
\ee 
The stellar mass bound to $m_2$ is effective in speeding the inspiraling
of $m_2$ only when $m_2$ is at large radii (e.g. $\ga10$--$100\pc$ in
Fig.~\ref{fig:timefric} below).

When $m_2$ sinks to the radius at which $M_*(r)\approx M_\bullet$
(e.g. $r\sim1$--$10a\H$ in Fig.~\ref{fig:Mr} below),
the two BHs form a ``bound'' binary system (this marks the transition from the
dynamical friction stage to the non-hard binary stage).
(Here, the definition of ``bound'' is somewhat arbitrary; we only
intend to mean that the gravitational force on $m_2$ is dominated by
$m_1$ rather than by the stars.)
Though the BBH orbital decay in the non-hard binary stage
comes from both dynamical friction from distant stars and three-body
interactions between the BBH and the stars passing the BBH vicinity,
we still use equation (\ref{eq:frictime}) to estimate the BBH evolution
timescales in both the dynamical friction stage and the non-hard binary stage
because scattering experiments with the restricted three-body approximation
basically give a similar hardening timescale to equation (\ref{eq:frictime})
[see equations 2, 16 and 18 in Quinlan (1996)].

Before the BBH becomes bound [$M_*(r)\ga M_\bullet$], the removed mass from
the galactic core can be ignored since both the mass and binding energy of the
smaller BH are less than the total mass and kinetic energy of the core stars
interior to the BBH orbit; thus the inspiral has negligible effect on the
dynamics of the core stars.
But, during the non-hard binary stage, depletion of the stars that interact
with the BBH can be significant in some galaxies.
The mass removing rate from the galactic core caused by interactions between
stars and one or both of the BHs, $J_{\rm rm}(a)$, is defined by
\be
J_{\rm rm}(a)\equiv\frac{1}{M_\bullet}\frac{\d M_{\rm rm}}{\d\ln(1/a)},
\label{eq:Jrm}
\ee
where $M_{\rm rm}$ is the stellar mass removed from the galactic core.
In \S~\ref{sec:lr}, we introduced the concept of loss region 
(loss cone for spherical galaxies and loss wedge for axisymmetric galaxies)
for a hard BBH, but equations (\ref{eq:Mlc}), (\ref{eq:Mlw}) and (\ref{eq:Mlr})
(i.e., the total stellar mass in the loss region) can be generalized to the
non-hard binary stage to reflect the mass of the stars which may have close
encounters with the BBH.
Here, the stars which may have close encounters with the BBH can be either the
original low-angular momentum stellar population or those precessed from
high-angular momentum orbits by tidal forces.
Before the BBH becomes hard, if the mass of the initial stellar population in
the loss region is not enough to provide the removed stellar mass from the
galactic core $M_{\rm rm}$ (see equation \ref{eq:Jrm}),
the BBH would lose energy mainly by interactions with distant stars,
which becomes more and more inefficient as the BBH hardens,
and by three-body interactions with the low-angular momentum stars which are
diffused from high-angular momentum orbits by two-body relaxation, which
will dominate the BBH hardening timescales in the hard binary stage (see
equations \ref{eq:th}, \ref{eq:Flc} and \ref{eq:Flw}).
Thus, the BBH hardening timescale should be higher than those estimated from
equations (\ref{eq:fricforce})--(\ref{eq:frictime}) in the non-hard binary
stage, and smoothly increase to connect the hardening timescale in the hard
binary stage.
Here, we use the linear approximation [for the $\ln(t\h)$--$\ln(a)$ relation]
to obtain the BBH hardening timescale at $a\h\le a\le a_{\rm dp}$:
\be
\ln[t\h(a)]=\ln[t\h(a\h)]+\ln[t\df(a_{\rm dp})/t\h(a\h)]\ln(a/a\h)/\ln(a_{\rm
dp}/a\h), 
\label{eq:thdf}
\ee
where $a_{\rm dp}$ is the semimajor axis where the initial population of stars
in the loss region are all removed from the loss region
(which will be determined in \S~\ref{sec:Mlr}), $t\h(a\h)$ is obtained
from \S~\ref{sec:evlafh} and $t\df(a)$ is obtained from equations
(\ref{eq:fricforce})--(\ref{eq:frictime}).
Considering the possible BBH energy loss caused by gravitational radiation,
the BBH evolution timescale $t_{\rm evol}(a)$ is given as
equation (\ref{eq:tevol}) and $t\h(a)=t_{\rm gr}(a)$ may happen before the
BBH becomes hard.

\section{Results}\label{sec:res}
\subsection{Galaxy properties}\label{sec:galpro} 

\noindent
In the past several years, images from the {\it Hubble Space Telescope} (HST)
have revealed many details about the central regions of nearby galaxies, with
a resolution of 0.1\arcsec, corresponding to distances of several pc or
$10^5$--$10^6(M_\bullet/10^8\msun$) Schwarzschild radii for typical target
galaxies \cite{Betal96}.  The inner surface brightness profiles of the
galaxies are well fitted with a five-parameter fitting function -- the Nuker
law: 
\be
I(r)=2^{\frac{\beta-\gamma}{\alpha}}I\b\left(\frac{r}{r\b}\right)^{-\gamma}
\left[1+\left(\frac{r}{r\b}\right)^\alpha\right]^
{-\frac{\beta-\gamma}{\alpha}}.
\label{eq:nukerlaw}
\ee
The asymptotic logarithmic slope inside $r\b$ is $-\gamma$, the asymptotic
outer slope is $-\beta$, and the parameter $\alpha$ parameterizes
the sharpness of the break. The break radius $r\b$ is the point of maximum
curvature in log-log coordinates. The ``break surface brightness'' $I\b$ is the
surface brightness at $r\b$.
Elliptical galaxies and spiral bulges (hot galaxies) can be
classified into two main types according to their inner surface brightness
profiles ($I\propto r^{-\gamma}$ when $r\rightarrow 0$).
Hot galaxies with a quite shallow inner power-law slope ($\gamma\la 0.3$) are
classified as ``core'' galaxies and those with a steep slope ($\gamma\ga 0.5$)
are labelled as ``power-law'' galaxies.
The difference between core and power-law profiles has no direct connection
with the presence of a disk, whether seen edge-on or face-on, although
power-law galaxies have disky isophotes \cite{Fetal97}.
Core galaxies are luminous ($M_{\rm V}<-20.5$) with large central BHs and
power-law galaxies are faint ($M_{\rm V}>-22$) with small central BHs.
Early discussions of the demography of the central BHs focused on the
correlation of BH mass with galaxy luminosity.
It is now believed that BH mass is correlated more tightly with galactic
velocity dispersion, which suggests a causal connection between the
formation and evolution of the BH and the bulge \cite{Getal00,FM00}. 

We shall apply the theoretical results of the preceding sections to estimate
what the evolution timescales of the BHs in the centers of nearby galaxies
would be if they were binary.

We use the HST sample of hot galaxies compiled in the paper by Faber et al.
(1997). Among the 61 galaxies in that paper, 41 galaxies (Table 1) have values
for the Nuker law parameters, the stellar mass-to-light ratio $\Upsilon$
(constant for each galaxy) and the half-light radius, where $\Upsilon$ is
determined by normalizing to the central velocity dispersion based on spherical
and isotropic models fitted to the Nuker-law profile.
The isotropized DF $\iso f(\E)$ of the galaxies (see equation \ref{eq:nuf})
can be obtained by the Eddington formula \cite{BT87}: 
\be
\iso f(\E)\approx\frac{1}{\sqrt{8}\pi^2}\frac{\d}{\d\E}\int^{\E}_{-\infty}
\frac{1}{m_*}\frac{\d\rho}{\d\psi}\frac{\d\psi}{\sqrt{\psi-\E}},
\label{eq:Eddington}
\ee
where
\be
\rho(r)=\Upsilon j(r)=-\frac{\Upsilon}{\pi}\int_r^\infty\frac{\d I}{\d R} \frac{\d R}{\sqrt{R^2-r^2}}.
\ee
The total mass of the central BHs is estimated by:
\be
M_\bullet=1.2(\pm 0.2)\times10^8\msun\left(\frac{\sigma\e}{200\kms}\right)^{3.75(\pm 0.3)},
\label{eq:msigma}
\ee
where $\sigma\e$ is the luminosity-weighted line-of-sight velocity dispersion
inside the half-light radius \cite{Getal00}.
The BH mass--velocity dispersion relation due to Merritt \& Ferrarese (2001)
uses central velocity dispersion and the velocity dispersion has a different
exponent, 4.72$(\pm0.36)$. In this paper, we use equation (\ref{eq:msigma})
to estimate the central BH mass.
The results will not be affected much if we equate the central
velocity dispersion to $\sigma\e$ and use the different exponent 4.72.
Of these 41 galaxies in Table 1, 9 DFs obtained from equation
(\ref{eq:Eddington}) are negative, perhaps because the spherical and isotropic
dynamical model is not appropriate or because the Nuker law is a poor fit,
and these are deleted from our sample.
All of these 9 galaxies are core galaxies with shallow central cusps
($0\la\gamma\la 0.04$).
Generally, $\sigma\e$ is determined by the gravitational potential of stars
and affected little by the gravitational potential of the central BH, but that
is not true for two galaxies with $\gamma>1$, which are also deleted from
our sample. Thus, a total of 30 galaxies is left in our study.

\subsection{Total stellar mass in the full loss region}\label{sec:Mlr}

\begin{figure}
\centerline{\psfig{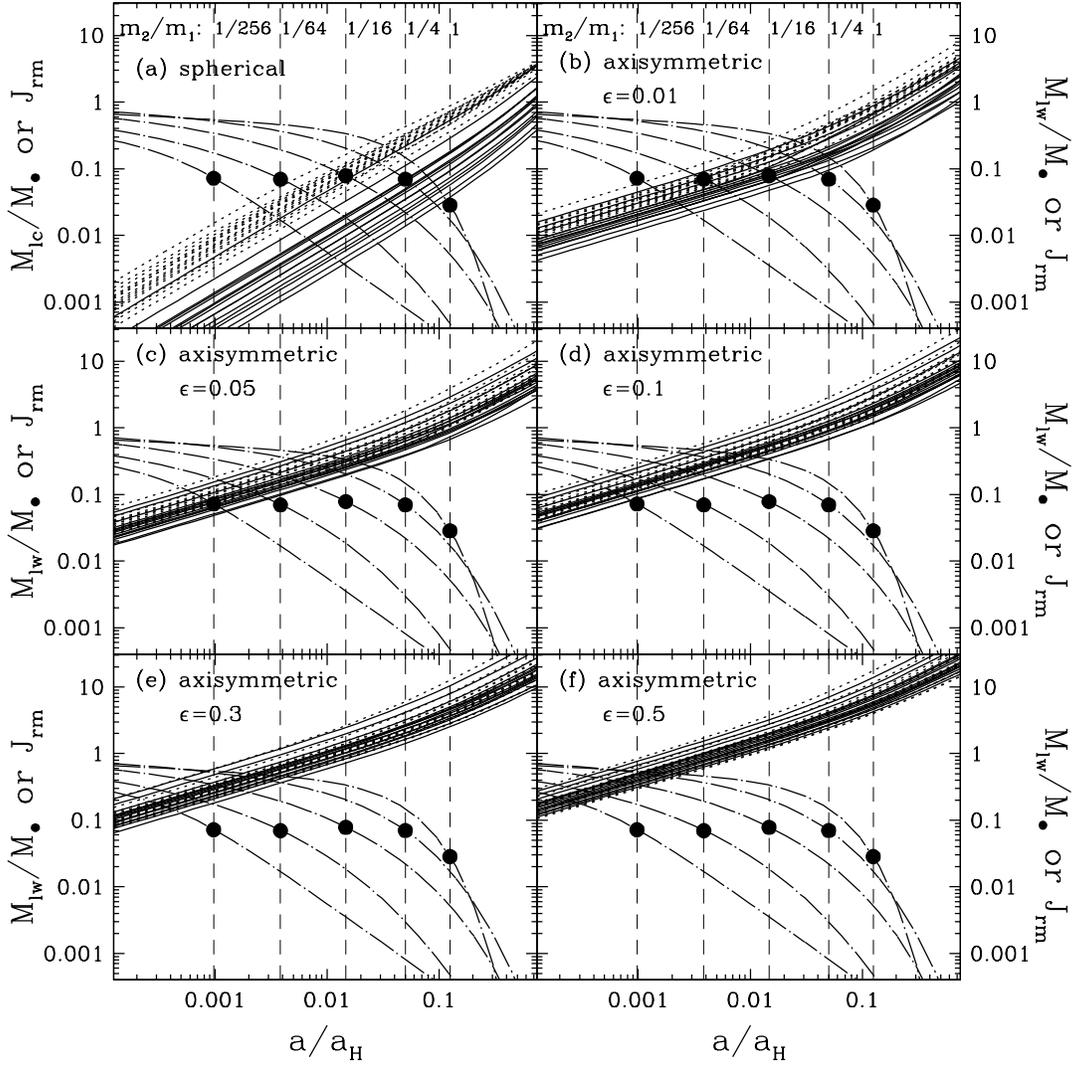}}
\caption{The total stellar mass in the full loss cone/wedge
(for spherical/axisymmetric galaxies) and the removing rate from the
galactic core as a function of $a/a\H$ ($a\H$ represents the radius of the
sphere of influence of the BH $M_\bullet$, cf. equation \ref{eq:aH}).
The solid lines (for core galaxies in Table 1) and the dotted lines (for
power-law galaxies in Table 1) represent the total stellar mass in the full
cones $M\lc(a)$ in panel (a)
(cf. equation \ref{eq:Mlc}) and the total stellar mass in the full loss
wedges $M\lw(a)$ in panels (b)--(f) (cf. equation \ref{eq:Mlw}).
The dot-dashed lines represent the removing rate from the galactic core
$J_{\rm rm}$ (see equation \ref{eq:Jrm}) for the five different $m_2/m_1$ ratios
($m_2/m_1=1/256, 1/64, 1/16, 1/4, 1$) which can be obtained from
Quinlan's simulation (1996).
The vertical dashed lines represent the place where the BBH first becomes hard
and the corresponding $m_2/m_1$ ratio is labelled at the right side of each
dashed line.
For each given $m_2/m_1$ ratio, the crossing of the vertical dashed line
($a=a\h$) and the removing rate curve, $J_{\rm rm}$, is labelled as a solid
circle.
The total mass in the full loss region increases with increasing $\epsilon$.
The $M\lc(a\h)$ and $M\lw(a\h)$ also increase with increasing $m_2/m_1$ ratios.
For the cases in which the solid or dotted lines are higher than the solid
circles at $a=a\h$, the removed stellar mass from galactic cores during the
non-hard binary stage can be ignored and the loss region is approximately full
at $a=a\h$ (see details in \S~\ref{sec:Mlr}).
}
\label{fig:mlrjrm}
\end{figure}

Before studying the details of the BBH evolution timescales, we will first
generalize equations (\ref{eq:Mlc}) and (\ref{eq:Mlw}) to the non-hard binary
stage to obtain the total stellar mass in the full loss cone/wedge (for
spherical/axisymmetric galaxies) to see if the total stellar mass in the full
loss cone/wedge is large enough to provide the removed stellar mass from
galactic core before the BBH becomes hard (Fig.~\ref{fig:mlrjrm}).

Quinlan's simulation (1996) gives the mass removing rate from the galactic
core caused by interactions between a BBH and stars ($J_{\rm rm}$,
see equation \ref{eq:Jrm}) as a function of $\sigma\c/v\c$ (or $a/a\h$) for
five different BH mass ratios ($m_2/m_1$=1/256, 1/64, 1/16, 1/4, 1), where
$v\c\equiv\sqrt{GM_\bullet/a}$.
We can roughly obtain the removing rates for other BH mass ratios by
interpolating the rates of those five mass ratios.
As shown in Figure~\ref{fig:mlrjrm} (or see Fig.~5 in Quinlan 1996),
the removing rate $J_{\rm rm}$ is small at large semimajor axes
(e.g. $J_{\rm rm}\la10^{-3}$ at $a/a\H\simeq0.3$)
and increases steeply as the BBH hardens, and we may approximate equation
(\ref{eq:Jrm}) as $M_{\rm rm}(a)\simeq M_\bullet J_{\rm rm}(a)$.
The removing rate is about 0.03--0.1 when the BBH becomes hard,
and then increases towards a constant in the range 0.1--1 as the BBH hardens.
This is roughly consistent with the factor $1/(2K)\sim0.3$
($Kf_a=1.56$, $f_a\sim 1$) in equation (\ref{eq:th})
or the energy gain in equation (\ref{eq:DeltaE})
(i.e., soon after the BBH becomes hard, especially when $a\ll a\h$,
most stars having close encounters with the BBH will be removed from the
galactic core).

Figure~\ref{fig:mlrjrm}(a) shows the total stellar mass in the full loss cone
$M\lc(a)$ and the removing rate as a function of $a$ if the galaxies in Table
1 are spherical. 
As seen from Figure~\ref{fig:mlrjrm}(a), with the decrease of $a$,
$M\lc(a)$ decreases and the mass removing rate
from the galactic core $J_{\rm rm}(a)$ increases.
For the BBH with equal masses ($m_2/m_1=1$), we have
$M\lc/M_{\rm rm}\simeq M\lc/M_\bullet J_{\rm rm}\ga 1$ at $a\ga a\h$ [i.e., in
Fig.~\ref{fig:mlrjrm}a, the solid circle representing the stellar removing
rate $J_{\rm rm}(a=a\h)$ for $m_2/m_1=1$ is located below all the $M\lc(a\ga
a\h)/M_\bullet$ curves]; so the mass of the removed stars from the galactic
core during the non-hard binary stage can be ignored and the loss cone is
approximately full at $a=a\h$.  Compared with the BBH with equal masses,
the BBH with low-$m_2/m_1$ ratio becomes hard at smaller semimajor axis and
the removed mass may be larger than the mass in the full loss cone before the
hard binary stage [i.e., in Fig.~\ref{fig:mlrjrm}a, solid circles for
low-$m_2/m_1$ ratios are located above some of the $M\lc(a\ga a\h)/M_\bullet$
curves].
For the BBHs whose loss cone is depleted before the hard binary stage begins
(e.g. 20 out of our sample of 30 galaxies for $m_2/m_1=1/16$),
we obtain the depletion semimajor axes $a_{\rm dp}$ approximately by setting
$M\lc(a_{\rm dp})/M_\bullet=J_{\rm rm}(a_{\rm dp})$ and we will use equation
(\ref{eq:thdf}) to obtain the BBH hardening timescales at $a\h<a<a_{\rm dp}$
in \S~\ref{sec:ressph}. 
Note that in Figure~\ref{fig:mlrjrm}(a), at a given
$a$($\la 0.1a\H$), $J_{\rm rm}(a)$ decreases with BH mass ratios; and
hence $a_{\rm dp}$ also decreases with BH mass ratios.

Figure~\ref{fig:mlrjrm}(b)--(f) shows the total mass in the full loss wedge
$M\lw(a)$ and the removing rate as a function of $a$ if the galaxies in
Table 1 are axisymmetric.
The total mass in the full loss wedge depends on the degree of flattening
$\epsilon$ (defined in \S~\ref{sec:Faxis}), which is assumed to be a constant
here.
The average mass ellipticity of the sample in our study, which can be obtained
from Table 2 in the paper by Faber et al. (1997), is about 0.26.
Their average $\epsilon$ is $\sim0.36$ if equation (\ref{eq:epsilonb}) is
applied.
As seen from Figure~\ref{fig:mlrjrm}(b)--(f) ($\epsilon$=0.01, 0.05, 0.1, 0.3 and
0.5), $M\lw(a)$ increases with the increase of $\epsilon$.
If $\epsilon$ is very small (e.g. $\epsilon<0.01$), the loss wedge is
approximately full at $a=a\h$ only for $m_2/m_1\simeq1$ as in the spherical
case.
When $\epsilon\ga0.1$, the loss wedge is approximately full at $a=a\h$ for
$m_2/m_1\ga1/256$ (i.e., the corresponding solid circle is located below all
the $M\lw/M_\bullet$ curves at $a\ga a\h$ in Fig.~\ref{fig:mlrjrm}d--f).

For triaxial galaxies, the maximum mass of the stars which may have close
encounters with the BBH is not less than the mass in the full loss
wedge of axisymmetric galaxies with the same parameter $\epsilon$. 
In the study below of the evolution of BBHs in triaxial galaxies (see
\S~\ref{sec:restri} below), we will mainly consider BBHs with equal masses.
In this case, the removed stellar mass during the non-hard binary stage
can be ignored and the loss region is approximately full at $a=a\h$ as in
axisymmetric galaxies (cf. Fig.~\ref{fig:mlrjrm}b--f).  For other cases in
triaxial galaxies (varying BH mass ratios at a given $\epsilon$), the
variation of the BBH evolution should follow similar trends to those in
axisymmetric galaxies, and we need not consider the maximum mass of the stars
which may have close encounters with the BBH in triaxial galaxies.

\subsection{Spherical galaxies}\label{sec:ressph}

\begin{figure}
\centerline{\psfig{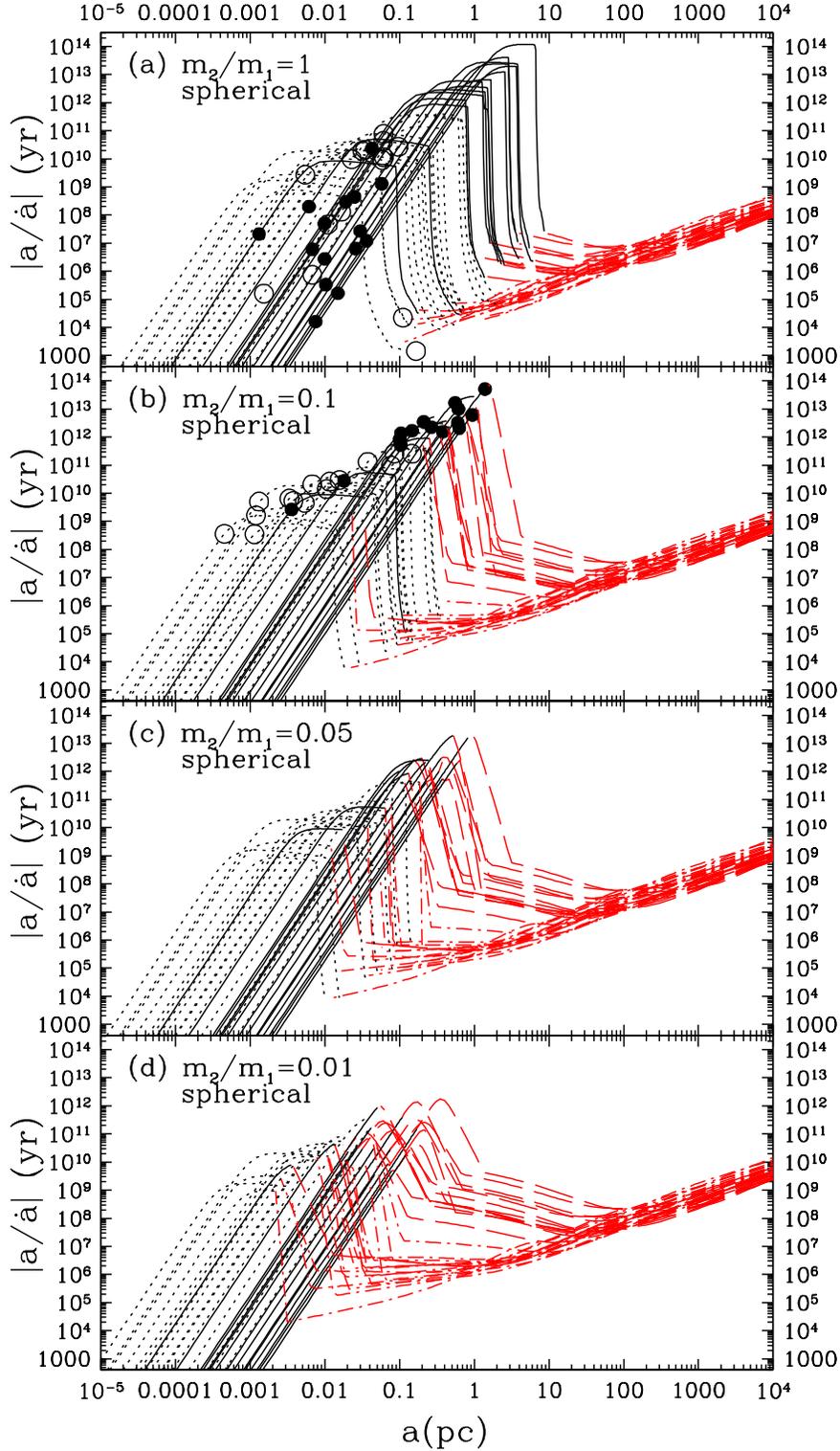}}
\caption{BBH evolution timescales for spherical galaxies.
The panels show different mass ratios $m_2/m_1$ and
the curves represent timescales $|a/\dot a|$ for the galaxies in Table 1.
The long dashed lines (for core galaxies: $\gamma\simlt 0.3$) and the dot-short
dashed lines (for power-law galaxies: $\gamma\simgt 0.5$) represent the
timescales before the hard binary stage;
and the solid lines (for core galaxies) and the dotted lines (for power-law
galaxies) represent the timescales after the BBHs become hard.
In panel (a), the solid circles (for core galaxies) and the open circles
(for power-law galaxies) represent the maximum Brownian motion magnitudes of
the BBH centers of mass ($r_{\rm ran}$ in equation \ref{eq:rran}).
In panel (b), the solid circles (for core galaxies) and the open circles
(for power-law galaxies) mark the boundary between the hard binary and
gravitational radiation stage or the place where gravitational radiation
becomes dominant in the non-hard binary stage
(i.e., $a=a_{\rm gr}$ when $t\h=t_{\rm gr}$).
BBHs in power-law galaxies have shorter lifetimes than those in core galaxies.
With decreasing mass ratios, the evolution timescales at large radii
(e.g. $a\simgt 10^2\pc$) increase and hence BBHs with very low mass ratios
(say, $\simlt 0.001$) cannot be formed by mergers of galaxies.
The bottleneck stages shift to smaller radii as the mass ratio
decreases.
In most core galaxies, the bottleneck timescales become shorter as the
mass ratio decreases; while in most power-law galaxies, they are affected
little by mass ratios.
The lifetime of BBHs is not affected by their Brownian motion.
See details in \S~\ref{sec:ressph}.
}
\label{fig:timesph}
\end{figure}

\noindent
Assuming that the galaxies in Table 1 are spherical, we obtain the BBH
evolution timescales as a function of BBH semimajor axis (or the separation
of the two BHs before they become bound) and the BH mass ratio $m_2/m_1\le1$.
The results are shown in Figure~\ref{fig:timesph}, where the BBH evolution
curves are composed of two parts: one comes from equations
(\ref{eq:th})--(\ref{eq:tevol}) and (\ref{eq:Fdrainsph})--(\ref{eq:Fsph}) in
\S~\ref{sec:evlafh} describing the evolution in the hard binary stage and the
gravitational radiation stage (dotted or solid lines), and the other from
equations (\ref{eq:fricforce})--(\ref{eq:mboundratio}) or equation
(\ref{eq:thdf}) in \S~\ref{sec:evlbfh} describing the evolution in the
dynamical friction stage and the non-hard binary stage (long dashed or
dot-short dashed lines).
A characteristic feature of all these plots is a pronounced peak in the
evolution timescale between 0.001 and 10 pc, where dynamical friction and
gravitational radiation are both relatively ineffective. This is always the
slowest evolution stage if we do not consider the possibly longer timescales
at very large radii, $a\ga 10^4\pc$. We shall call this stage the
``bottleneck''.

To study the effect of different BH mass ratios on the BBH evolution,
we will first consider BBHs with equal masses (in \S~\ref{sec:resspheqms}),
and then consider those with unequal masses (in \S~\ref{sec:ressphueqms}),
though actually there is no strict dividing line between these cases.

\subsubsection{Evolution of BBHs with equal BH masses}\label{sec:resspheqms}

\noindent
In this subsection, we will consider the evolution of BBHs with equal BH masses
($m_2/m_1=1$, Fig.~\ref{fig:timesph}a) for which the removed stellar mass from
the core during the non-hard binary stage can be ignored and the loss cone is
approximately full at $a=a\h$ [$\eta(\E)\simeq 1$ in equation \ref{eq:Mlc}]
(see \S~\ref{sec:Mlr} or Fig.~\ref{fig:mlrjrm}a).  

As seen from Figure~\ref{fig:timesph}(a), the overall evolution of the BBHs has
the following trends: starting at large radii ($\sim 10\kpc$), the evolution
timescales first decrease with decreasing radii;
then increase at some intermediate radii to reach the bottleneck;
and finally when the BBH semimajor axis
$a$ is small enough ($\la 10^{-3}$--1$\pc$), the evolution timescales decrease
with decreasing $a$.
The bottleneck occurs during the hard binary stage.

\begin{figure}
\centerline{\psfig{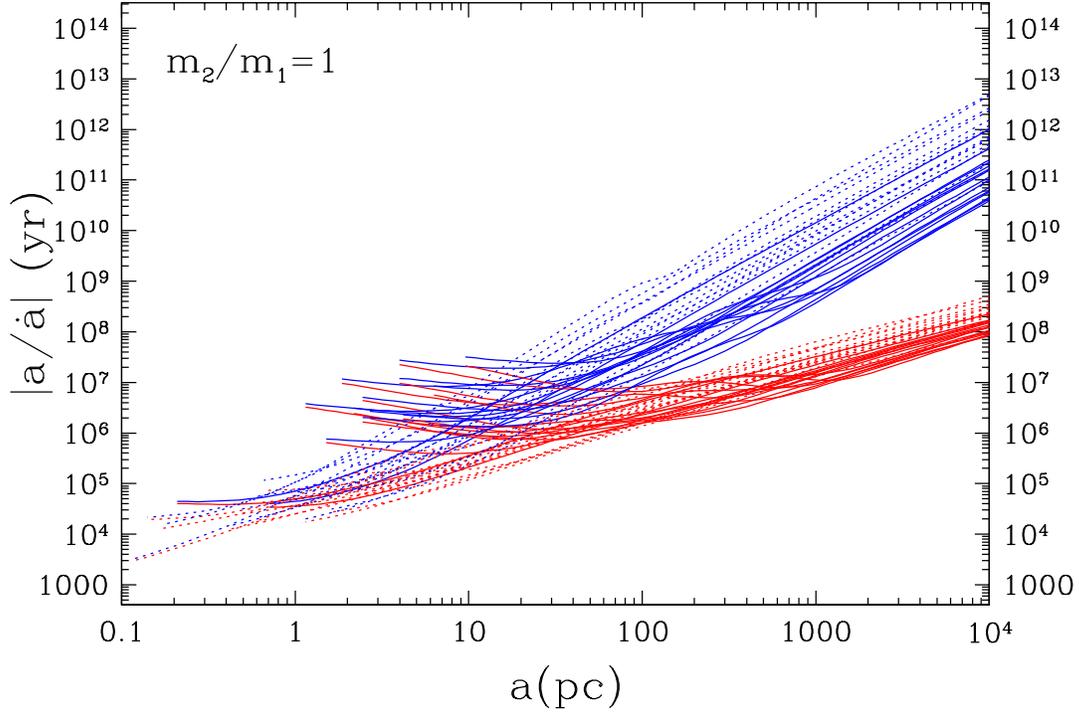}}
\caption{Evolution timescales of BBHs (with equal BH masses) as a function of
the BBH separation or the orbit semimajor axes in the dynamical
friction stage and non-hard binary stage.
The solid lines represent core galaxies and the dotted lines represent
power-law galaxies.
At small semi-major axis, the curves are cut off at $a=a\h$.
The curves are composed of two groups, which is easy to distinguish at large
radii ($a\sim 10^3-10^4\pc$).
The lower group ($|a/\dot a|\sim 10^{8}$--$10^{9}\yr$ at
$a\simeq10^4\pc$) are the evolution timescales including the bound stars
around $m_2$;
and the higher group ($|a/\dot a|\sim 5\times10^{10}$--$5\times10^{12}\yr$ at
$a\simeq10^4\pc$) are the evolution timescales without including the
bound stars around $m_2$ (see \S~\ref{sec:evlbfh}).
At small radii ($a\rightarrow a\h$), the two groups have almost the same
timescales.
At large radii ($a\simgt 100\pc$), the evolution timescales decrease
with decreasing $a$; at small radii ($a\simlt 100\pc$), the evolution
timescales increase for most core galaxies and decrease for most
power-law galaxies (see \S~\ref{sec:resspheqms}).
The evolution timescales at large radii are shorter than the Hubble time
$10^{10}\yr$ if and only if we assume that the BHs are accompanied by the
tidally truncated remnant of stars from their original host galaxy.
}
\label{fig:timefric}
\end{figure}

Before the BBH becomes hard, the evolution timescales are always shorter than
the Hubble time $t\Hubble=10^{10}\yr$; these relatively short timescales
reflect the enhancement of dynamical friction due to the bound stellar mass
around $m_2$.
Figure~\ref{fig:timefric} shows that without including the bound stellar mass
around $m_2$, the dynamical friction timescales would be much larger than
$t\Hubble$ at large radii
(e.g. $\sim 5\times10^{10}$--$5\times10^{12}\yr$ at $10\kpc$).
The dynamical friction timescale decreases with the decrease of $r$, and when
$r$ decreases to the place where $m_1$ dominates the potential, we have
$|r/\dot r|\propto v_{m_2}^3(r)/\rho(r)\propto r^{-1.5}/r^{-\gamma-1}\propto
r^{\gamma-0.5}$ (see equations \ref{eq:fricforce}--\ref{eq:frictime}).
Thus, at small radii ($a\la100\pc$), with the decrease of $r$, the evolution
timescale continues decreasing for power-law galaxies ($\gamma\ga 0.5$),
but increases for core galaxies ($\gamma\la 0.3$).

\begin{figure}
\centerline{\psfig{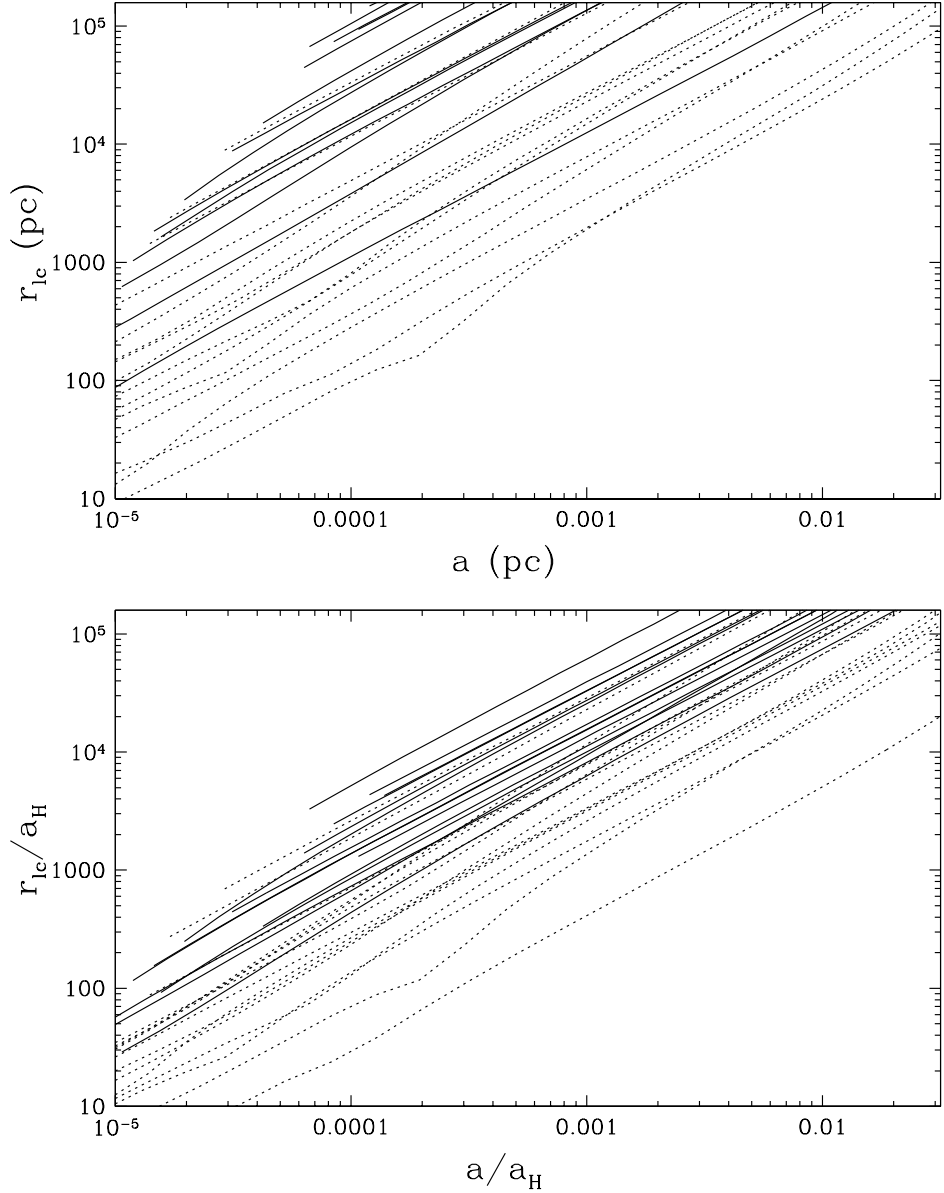}}
\caption{The semimajor axes $a$ of the BBHs versus
the transition radii $r\lc$ between the pinhole ($r>r\lc$) and the
diffusion ($r<r\lc$) regimes in spherical galaxies.
The lower ends of the lines are cut off at the Schwarzschild radius
$r_{\rm s}=2GM_\bullet/c^2$.
The solid and dotted lines have the same meaning as those in
Figure~\ref{fig:timesph}.
At a given $a$, most core galaxies have larger transition radii
$r\lc$ than power-law galaxies because core galaxies generally have
more massive BHs and lower central densities.
In the hard binary stage, if the loss cone in spherical galaxies is depleted,
the inner parts of the galaxies are generally in the diffusion regime.
}
\label{fig:rlca}
\end{figure}

\begin{figure}
\centerline{\psfig{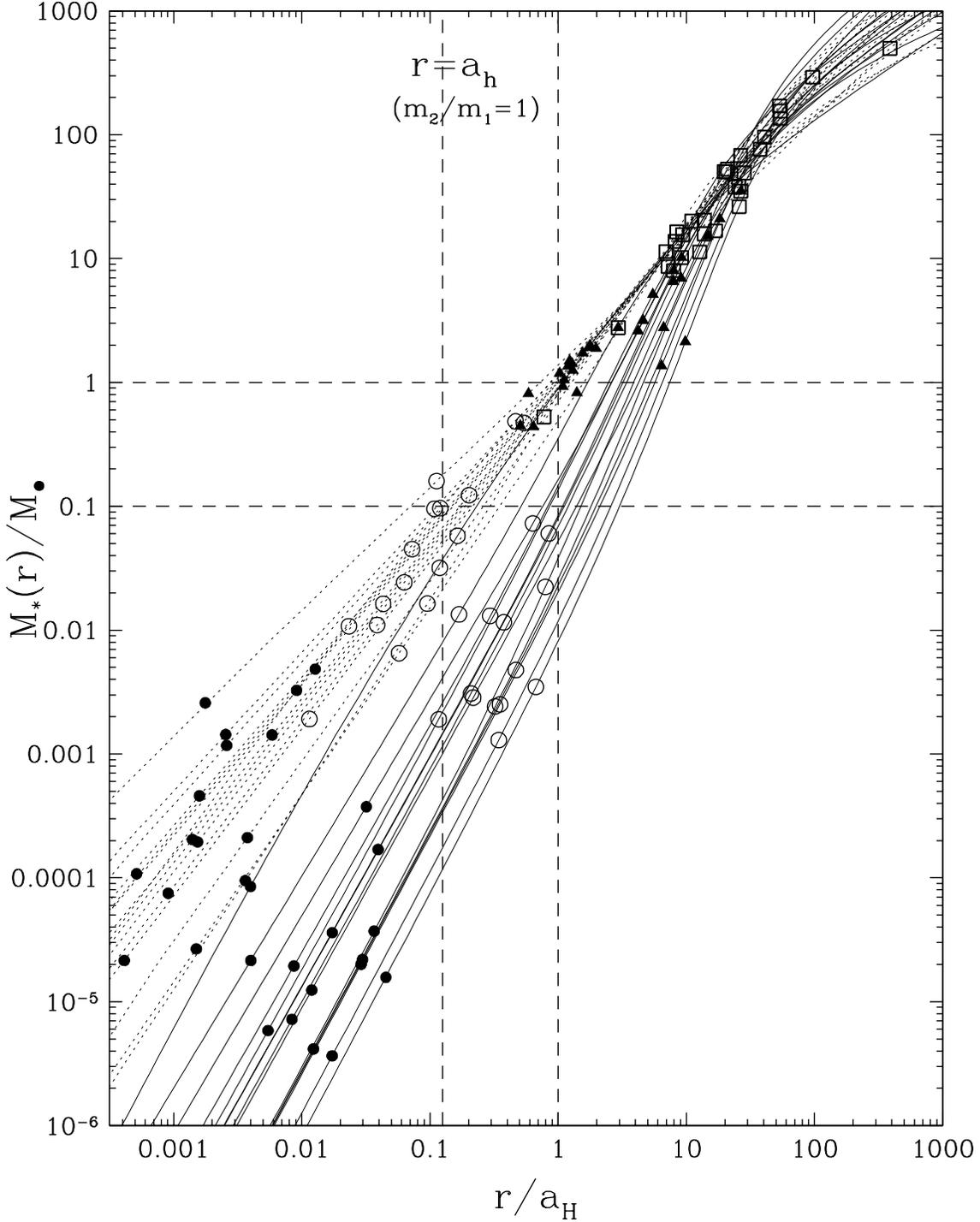}}
\caption{
The ratios of enclosed stellar mass within radius $r$ to BBH total mass.
The solid and dotted lines have the same meaning as those in
Figure~\ref{fig:timesph}.
The left vertical dashed line represents the semimajor axis $a\h$ where the BBH
with equal masses first becomes hard.
The open squares represent the break radii of the surface brightness profile
$r\b$ (cf. equation~\ref{eq:nukerlaw});
the open circles represent the collision radii $r\coll$
(equation \ref{eq:rcoll});
the solid circles represent the boundary between the hard binary and
gravitational radiation stage $a_{\rm gr}$ when $m_2/m_1=1$
in Figure~\ref{fig:timesph}(a).
In the hard binary stage, if the loss cone in spherical galaxies is depleted,
most of the stars that diffuse into the loss cone have the energy
$\E\peak\equiv\psi(r\peak)$ and the apocenters of the
radial orbits, $r\peak$, are labelled as solid triangles.
For hard BBHs, we have $r\peak\gg\max(r\coll,a\h)\simgt f_a a$ ($f_a\simeq 1$),
which helps to justify the generalization of solution of the Fokker-Planck
equation in the tidal disruption context to the BBH systems.
}
\label{fig:Mr}
\end{figure}

As seen from Figure~\ref{fig:timesph}(a), after the BBH becomes hard,
starting at $a=a\h$, the BBH evolution timescale increases
steeply with decreasing separation $a$, then becomes nearly flat, and then
decreases as $a^4$ once the BBH reaches the gravitational radiation stage.
Most of the BBH evolution timescales at the bottleneck are longer than the
Hubble time.
In these galaxies we might expect to find BBHs with semimajor axes in the range
0.01--10~pc.
BBHs in the core galaxies generally reach the bottleneck at larger
semimajor axes ($a\h\sim$1--10pc) and need more time to reach the
gravitational radiation stage than those in the power-law galaxies. 
At the beginning of the hard binary stage, the BBH orbital decay rate is
controlled by the
draining rate of the stars initially in the loss cone and the BBH evolution
timescale increases steeply as the stars in the loss cone are depleted.
The loss cones are depleted (generally when $a\la 0.1a\H$) before the BBHs are
close enough for gravitational radiation to be effective.
Thereafter, the clearing rates are controlled by diffusion of stars into the
loss cone.
Figure~\ref{fig:rlca} shows that the inner parts of most of the galaxies are
in the diffusion regime ($r<r\lc$) before the gravitational radiation stage
begins (generally $a_{\rm gr}/a\h\ga 0.001$ in Figure~\ref{fig:Mr}), though the
transition radii $r\lc$ decrease with the shrinking of the loss cone.
As mentioned in \S~\ref{sec:Fsph}, in the diffusion regime, the flux of stars
into the loss cone is insensitive to $a$, and hence the BBH evolution timescales
controlled by the stellar diffusion are nearly constant.
As seen from Figure~\ref{fig:timesph}(a), the bottleneck timescales in
core galaxies are larger than those in power-law galaxies, which reflects
the fact that core galaxies have lower central densities (hence, the stellar
diffusion rates into the loss cone are slower) and more massive BHs
[hence, the average relative change in the BBH energy caused by interactions
with a star passing the BBH vicinity is smaller
($|\Delta E/E|=2Km_*/M_\bullet$)] (see equation \ref{eq:th}).
Gravitational radiation becomes dominant at larger semimajor axes in core
galaxies, because their BBH bottleneck timescales are longer and 
the gravitational radiation timescale (equation \ref{eq:tgr}) is smaller for
large BHs. 

\subsubsection{Evolution of BBHs with unequal BH masses}\label{sec:ressphueqms}

\noindent
For the BBH with unequal masses, as discussed in \S~\ref{sec:Mlr},
low-angular momentum stars may have been depleted before the BBH first becomes
hard because the BBH becomes hard at smaller semimajor axes and the total
stellar mass in the full loss cone decreases with the decrease of the
semimajor axes.
Figure ~\ref{fig:timesph}(b)--(d) shows the evolution of the BBH with unequal
masses ($m_2/m_1<1$).

As seen from Figure~\ref{fig:timesph}(b)--(d), roughly, the overall evolution
trend of the BBHs with unequal masses looks like those with equal masses;
but the bottleneck stages of the BBHs
with unequal masses can be either in the non-hard binary stage or in the
hard binary stage, and tend to occur at smaller semimajor axes.
The bottleneck timescales of most of the BBHs are still longer than the
Hubble time.
Generally, the bottleneck timescales decrease with decreasing mass ratios in
most of the core galaxies, but remain unchanged in most of the power-law
galaxies. 

Before the BBHs become hard, at large radii ($r\sim 10^2$--$10^4\pc$),
the BBH evolution timescales increase with decreasing
$m_2/m_1$, but most of them are still smaller than the Hubble time
$t\Hubble=10^{10}\yr$ (Fig.~\ref{fig:timesph}b--d).
But when $m_2/m_1\la 0.01$, the BBH evolution timescales in some of the
galaxies exceed the Hubble time $t\Hubble=10^{10}\yr$ at $r\simeq10\kpc$.
So, in BBHs with $m_2/m_1\la 0.01$,
the orbit of the smaller BH would not decay
to the galactic center from the outer edge of the galaxy within a Hubble time.

For the BBH with unequal masses, if the removed mass from the galactic core
during the non-hard binary stage can be ignored and the loss cone is
approximately full when the BBH becomes hard, the evolution of the BBH after
$a=a\h$ would be like that of the BBH with equal masses, which is first
controlled by the depletion of the loss cone (when the evolution timescale
increases sharply) and then controlled by the stellar diffusion (see flat parts
of the BBH evolution timescale curves in Fig.~\ref{fig:timesph}b--d) and
gravitational radiation.
The bottleneck stage of the BBH with unequal masses shifts to smaller
semimajor axes because it becomes hard at smaller semimajor axis.
Note that the flat part (controlled by stellar diffusion) of the BBH evolution
curve of each galaxy, if any, has almost the same height in all the panels of
Figure~\ref{fig:timesph}, which reflects the fact that the BH mass ratio does
not affect the hardening time $t\h$ (see equation \ref{eq:th}).

For the BBH with unequal masses, if the loss cone is depleted when the BBH
first becomes hard [$\eta(\E)\simeq0$ in equation \ref{eq:Mlc}], the following
BBH evolution will be directly controlled by the stellar diffusion to the loss
cone (see the flat part of the BBH evolution curves, mostly for power-law
galaxies) and then gravitational radiation, or directly controlled by
gravitational
radiation (no flat part in the BBH evolution curves, mostly for core galaxies).
Before the BBH becomes hard, we obtain the place $a_{\rm dp}$
where the low-angular momentum stars are depleted by setting $M\lc(a_{\rm
dp})=M_{\rm rm}\simeq M_\bullet J_{\rm rm}(a_{\rm dp})$ (see equation
\ref{eq:Jrm}).  And we use equation (\ref{eq:thdf}) (log-linear approximation)
to obtain the BBH hardening timescales from $a_{\rm dp}$ to
$a\h$ because the evolution at $a\h<a<a_{\rm dp}$ is an intermediate process
between the evolution controlled by dynamical friction and the evolution
controlled by three-body interactions with the stars that diffuse from
high-angular momentum orbits; then, we use equation (\ref{eq:tevol}) to obtain
the BBH evolution timescales. 
As seen from Figure~\ref{fig:timesph}(b)--(d), the evolution timescale curves of
these BBHs increase steeply from $a=a_{\rm dp}(>a\h)$ and smoothly connect to
the timescales in the hard binary stage.
If the BBH evolution is controlled by stellar diffusion to the loss cone when
it just becomes hard, its lifetime is not affected by mass ratios.
For some of the BBHs, gravitational radiation may become the dominant
dissipative force when $a>a\h$, and in the non-hard binary stage their
evolution timescales show a peak or maximum (i.e., the bottlenecks) at
$a=a_{\rm gr}$ where $t_{\rm gr}(a_{\rm gr})=t\h(a_{\rm gr})$.
With decreasing BH mass ratio, the merger timescale becomes shorter because
the gravitational timescale $t_{\rm gr}$ (equation \ref{eq:tgr}) decreases as
$a^4$, though at a given $a$, $t_{\rm gr}$
is inversely proportional to mass ratio $m_2/m_1$ (when $m_2/m_1\ll 1$).
Here, as the BH mass ratio decreases, the bottleneck stage shifts to
smaller semimajor axes because either $a\h$ or $a_{\rm dp}$ decreases.

\subsubsection{Sharp change of the BBH evolution timescale at $a=a\h$}\label{sec:shpinc}

\noindent
As seen from Figure~\ref{fig:timesph}(a), the BBH evolution timescale increases
sharply at $a=a\h$. This sharp increase arises because the loss cone is
depleted almost as soon as the BBH becomes hard.
A similar sharp increase starting at $a=a\h$ also exists in
Figure~\ref{fig:timesph}(b)--(d) for those galaxies
whose loss cones are approximately full when the BBH first becomes hard.
In this subsection, we will see that the coincidence between the
increase in the evolution timescale and $a\h$ is
caused by some simplification in our analysis; but even if more realistic
treatment is considered, the sharp increase would also start at some place
close to $a=a\h$, though not exactly at $a=a\h$, and our conclusions would
not be significantly affected.

Note that in the analysis in \S~\ref{sec:lr} and \S~\ref{sec:evlafh},
before the BBH becomes hard, the velocity dispersion of the low-angular
momentum stars not removed from the galactic core is assumed to be unaffected
by interactions with the BBH; and after the BBH becomes hard,
each star having had close encounters with the BBH is removed from the galactic
core with an energy gain.
In fact, the heating of the stars is a continuous physical process,
i.e., in the non-hard binary stage, the stars should also be heated as the BBH
hardens even though they are not removed from the galactic core.

From scattering experiments with the restricted three-body approximation,
the BBH energy change due to stellar encounters in the non-hard stage is
generally very small
[e.g. $|\Delta E(a=10a\h)|\simeq 0.03|\Delta E(a=a\h)|\simeq 0.1m_*\sigma\c^2$,
cf. fig.~1 in Quinlan (1996)] and decreases steeply with increasing $a$
(which is consistent with the steep decrease of the stellar removing rate
from the core with increasing $a$ at $a>a\h$ in Fig.~\ref{fig:mlrjrm});
only when $a$ is close to $a\h$ does the relative change in the BBH energy,
$\Delta E/E$, increase steeply towards a constant (independent of $E$)
as the BBH hardens [cf. fig.~1 in Quinlan (1996)].
Hence, in the non-hard binary stage, the energy increase of a star
occurs through a slow diffusion process.
If all the energy loss of the BBH is used to
heat the low-angular momentum stars, the heating would become significant after
$Gm_1m_2/2a\simeq 3M\lc(a)\sigma^2\c/2$, i.e., after
\be
a\simeq 5a\h\left({2m_1\over m_1+m_2}\right)^{1/2}
\left[{0.03M_\bullet\over M\lc(a\h)}\right]^{1/2}\la 5a\h,
\label{eq:aheat}
\ee
where $M\lc(a)\simeq M\lc(a\h)a/a\h$ for $a\ll a\H$ (cf. equation \ref{eq:Mlc})
and $M\lc(a\h)/M_\bullet\sim0.03$--1 for those galaxies whose loss cone are
approximately full at $a=a\h$ in Figure~\ref{fig:mlrjrm}(a).
Thus, the heating of the stars becomes significant only at some place
close to $a\h$ and the sharp increase of the BBH evolution timescales would
also start not far away from $a=a\h$.
The semi-major axis $a\h$ is also the point where
the typical energy gain of a scattered star is just large enough for it to
escape from the galactic core. Thus, we expect a sharp increase in the
evolution timescale near $a\simeq a\h$.

\subsubsection{Testing assumptions}\label{sec:testassmp}

\noindent
We can use the results obtained above to test some of the assumptions
that we have made in our analysis.

\begin{enumerate}

\item In the hard binary stage, if the loss cone is depleted, 
the BBH decay is controlled by the stellar diffusion into the loss cone.
We define $\E\peak$ as the energy at the peak of stellar diffusion rates
$F^{\rm lc}(\E,a)$ (see equation \ref{eq:Flc}); 
and $\E\peak$ is generally near or smaller than $\psi(a\H)$, not at $\E\lc(a)$
because it is the factor $F_{\rm max}(\E)$ that dominates the shape of
$F^{\rm lc}(\E,a)$ in equation (\ref{eq:Flc}) (see also MT).
As seen from Figure~\ref{fig:Mr}, most of the stars contributing to the
BBH orbital decay have energy $\E\peak\ll GM_\bullet/f_a a$ or
$r\peak\gg\max(r_{\rm coll},a\h)\ge f_a a$, where $r\peak$ is defined by
$\E\peak=\psi(r\peak)$ and $r_{\rm coll}$ (see equation \ref{eq:rcoll}) is the
collision radius,
which helps to justify the generalization of solution of the
Fokker-Planck equation in the tidal disruption context to the BBH systems.
Figure~\ref{fig:Mr} also shows that the
stellar mass within $r\peak$ is higher than 0.1$M_\bullet$ for all the
galaxies, which confirms that resonant relaxation is not important here.

\item In the hard binary stage, when stellar diffusion dominates the BBH orbital
decay, the diffusion timescale into the loss cone at $\E\sim\E\peak$ is given
by
\begin{eqnarray}
t_{\rm diff} & \sim & \frac{M_*(r\peak)}{\int F(\E,a)\d\E}\frac{J\lc^2(\E\peak,f_a a)}{J\c^2(\E\peak)} \cr
& \la & (0.06\sim0.6)\left(\frac{Kf}{1.56}\right)
\left(\frac{100\sim10}{r\peak/a\h}\right)\left(\frac{a}{a\h}\right)\left|\frac{a}{\dot a}\right|,
\label{eq:tdiff}
\end{eqnarray}
which is less than the evolution timescale $|a/\dot a|$. Thus our use of
time-independent solutions to the Fokker-Planck equation is justified.

\item With equation (\ref{eq:minact}), we may also obtain the total mass of
the stars strongly interacting with the BBHs during the hard binary stage,
which is less than the core mass as seen from Figure~\ref{fig:minact}
($m_2/m_1=1$) and is much less when $m_2/m_1<1$.
The removed mass from the galactic core during the non-hard binary stage is
also small ($\sim 0.1M_\bullet$), as seen from Figure~\ref{fig:mlrjrm}.
So, the approximation that we ignore the response of the galactic structure
to BBH evolution is reasonable.

\begin{figure}
\centerline{\psfig{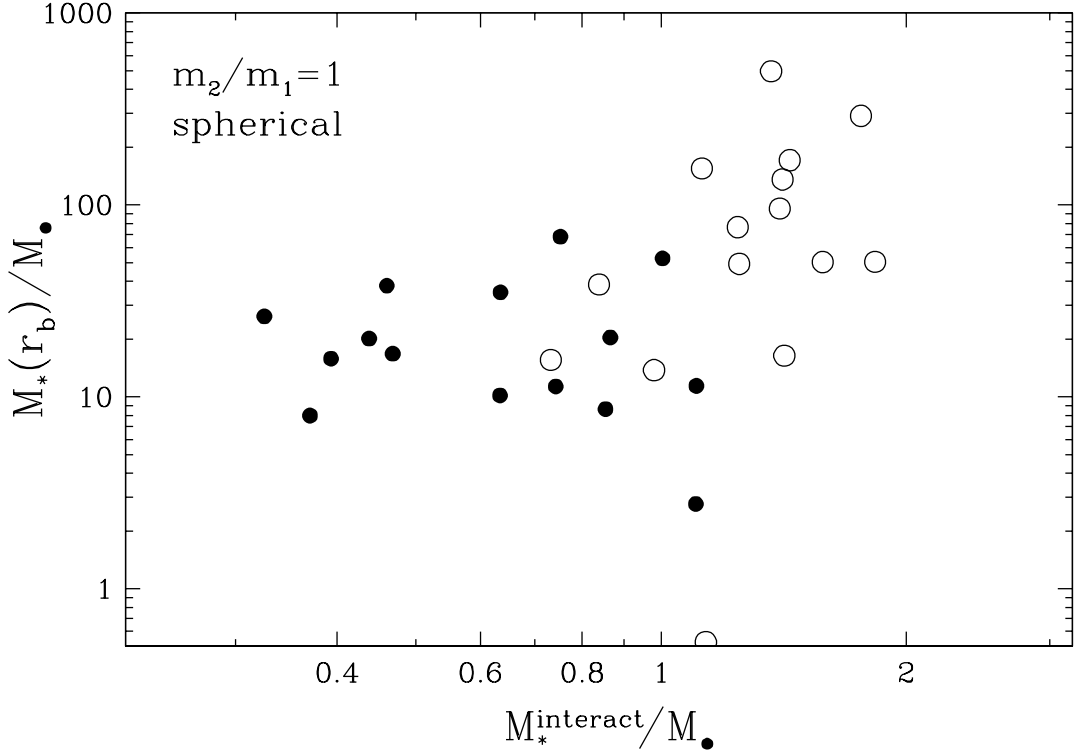}}
\caption{The core mass versus the total mass of stars interacting strongly with
BBHs ($m_2/m_1=1$, in spherical galaxies) during the hard binary stage
$a_{\rm gr}\le a\le a\h$ (see equation \ref{eq:minact} for $M_*^{\rm interact}$,
Fig.~\ref{fig:Mr} for $a\h/a_{\rm gr}$ or Fig.~\ref{fig:timesph}a for the BBH
evolution).
The solid circles represent core galaxies and the open circles represent
power-law galaxies.
For almost all of the galaxies, $M_*(r\b)>M_*^{\rm interact}$, which helps to
justify our neglect of the reaction of the core to the inspiral of the BHs.
}
\label{fig:minact}
\end{figure}

\item The center of mass of BBHs is not expected to be located exactly at the
galactic center \cite{BW76}.
The BBH center of mass should be in equipartition with the stars, which implies
that its rms radial excursion in core galaxies is
\be
r_{\rm ran}\sim \sqrt{\frac{\sigma^2_\bullet}{G\rho}}\sim \sqrt{\frac{m_*}{M_\bullet}}r\c=0.01\pc\sqrt{\frac{m_*}{\msun} \frac{10^8\msun}{M_\bullet}}\left(\frac{r\c}{100\pc}\right),
\label{eq:rran}
\ee 
where $\sigma_\bullet$ is the velocity dispersion of the BBH center of
mass.
The ``break radius'' $r\b$ is roughly equivalent to the core radius $r\c$ in
core galaxies. Power-law galaxies have no well-defined core, but replacing
$r\c$ with the break radius $r\b$ in equation (\ref{eq:rran}) should give an
upper limit to the likely motion of the BH. Hence, for both the core galaxies 
and the power-law galaxies, we set $r\c=r\b$ to obtain the wandering amplitude
of the center of mass of BBHs.
Wandering of the BBH center of mass might enlarge the loss cone, increase the
clearing rates and decrease the BBH hardening time
(e.g. Quinlan \& Hernquist 1997).  We
believe that this effect is generally not important in these spherical
galaxies for the following reasons: (1) For all the core galaxies,
gravitational radiation becomes dominant before their BBH ($0.01\la m_2/m_1\la
1$) semimajor axes decrease to the value $r_{\rm ran}$
(e.g. Fig.~\ref{fig:timesph}a).
(2) For most of the power-law galaxies, $a=r_{\rm ran}$ are located at the
flat parts of the evolution curves (see Fig.~\ref{fig:timesph}a) where stellar
diffusion dominates the BBH orbital decay, or $a=r_{\rm ran}$ are less than or
around the semimajor axes where the loss cones are depleted (e.g. for the BBHs
with lower mass ratios $0.01\la m_2/m_1<1$), and so the BBH lifetime is not
sensitive to the wandering of the BBH.
In Figure~\ref{fig:timesph}(a), for only two power-law galaxies, $a=r_{\rm ran}$
happens before their loss cones are depleted, but their BBH lifetimes are
shorter than the Hubble time even without considering the wandering of the BBHs.
(3) With decreasing $m_2/m_1$ ratios (say, $m_2/m_1\la 0.001$),
though $a=r_{\rm ran}$ may happen before the loss cone is depleted,
the dynamical friction timescales at large radii ($\ga10\kpc$) are longer
than the Hubble time and the smaller BH cannot sink into the center of the
galaxy with $m_1$.

\end{enumerate}

\subsection{Axisymmetric galaxies}\label{sec:resaxis}

\begin{figure}
\centerline{\psfig{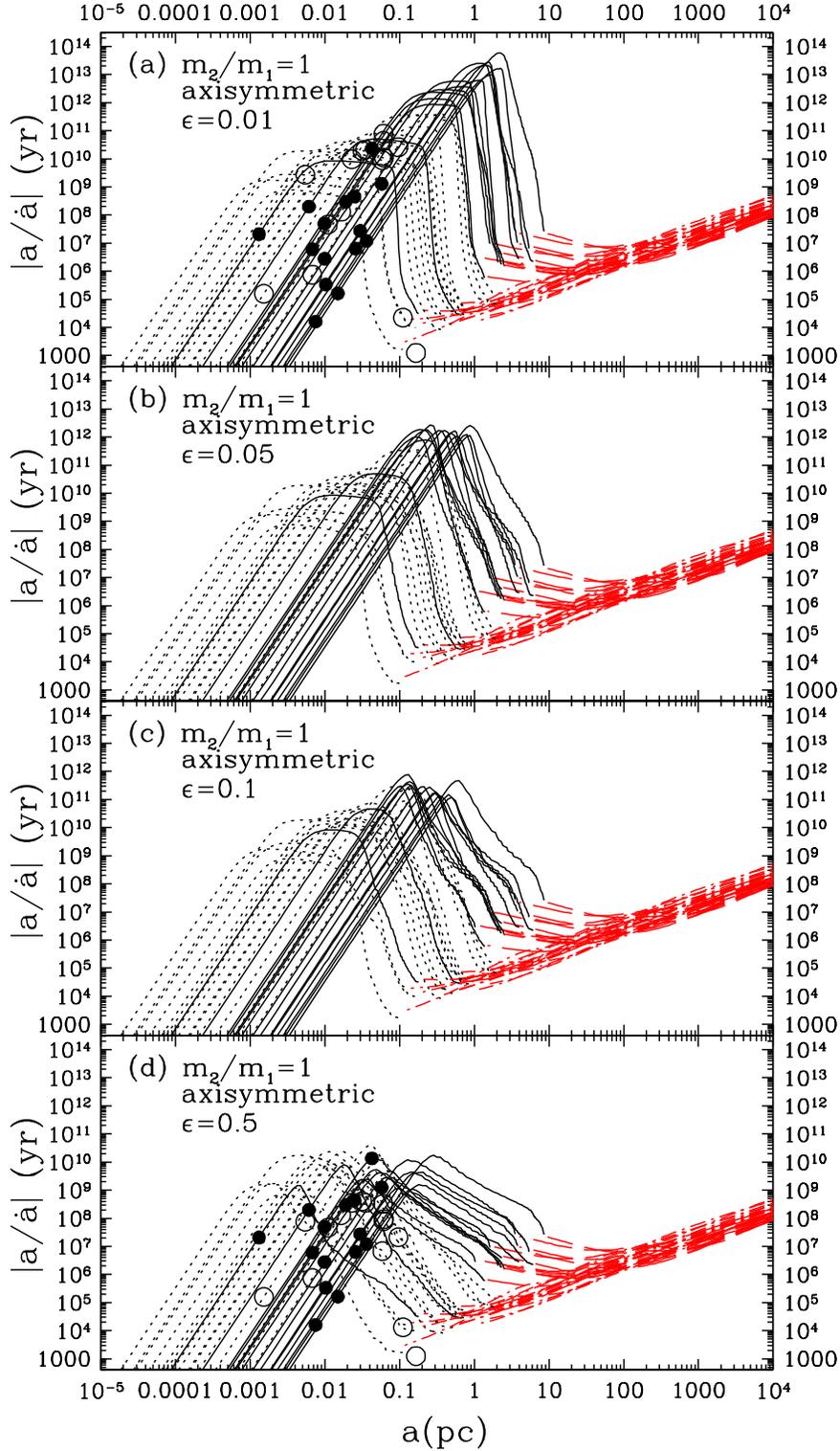}}
\caption{BBH evolution timescales for axisymmetric galaxies with $m_2/m_1=1$
and varying degrees of flattening $\epsilon$.
The curves have the same meanings as those in Figure~\ref{fig:timesph}.
In panels (a) and (d), the solid circles (for core galaxies) and the open
circles (for power-law galaxies) represent the maximum Brownian motion
magnitudes of the BBH centers of mass. The evolution curves before the
hard binary stage are the same as those in Figure~\ref{fig:timesph}.
The bottleneck is in the hard binary stage.
When $\epsilon$ is small (e.g. panel a), the BBH evolution reduces to the
spherical case (cf. Fig.~\ref{fig:timesph}a).
As $\epsilon$ increases, the bottlenecks shift to smaller semimajor axes
and many BBH lifetimes become shorter.
The Brownian motion of BBHs does not affect the BBH lifetime in nearly
spherical galaxies (e.g. panel a) or in core galaxies (e.g. panels a and d),
but may decrease the BBH lifetime in highly flattened power-law galaxies
(e.g. panel d).
See details in \S~\ref{sec:resaxis}.
}
\label{fig:timeaxis}
\end{figure}

\begin{figure}
\centerline{\psfig{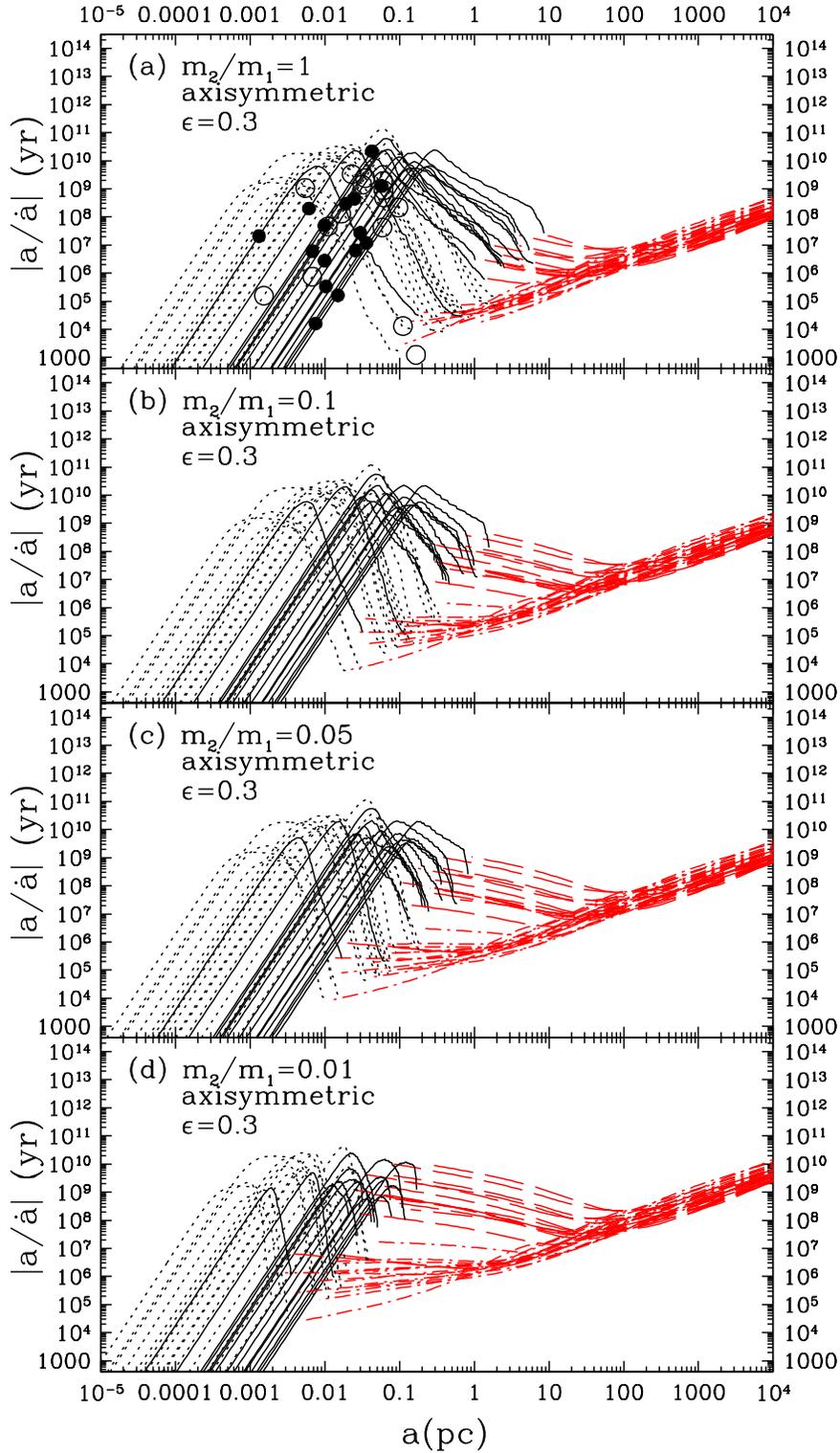}}
\caption{BBH evolution timescales for axisymmetric galaxies with
$\epsilon=0.3$ and varying mass ratios.
The curves have the same meanings as those in Figure~\ref{fig:timesph}.
In panel (a), the solid circles (for core galaxies) and the open
circles (for power-law galaxies) represent the maximum Brownian motion
magnitudes of the BBH centers of mass.
As the mass ratio decreases, the bottlenecks shift to slightly smaller radii
and the merger timescales become shorter.
Brownian motion of BBHs does not affect the BBH lifetime in core galaxies,
but may decrease the BBH lifetime in power-law galaxies.
}
\label{fig:timeaxis0.3}
\end{figure}

\noindent
Assuming that the galaxies in Table 1 are axisymmetric, we use equations
(\ref{eq:th})--(\ref{eq:tevol}) and (\ref{eq:TTmap})--(\ref{eq:Faxis}) to
obtain the BBH evolution timescales as a function of BBH semimajor axes after
the BBH becomes hard (equations \ref{eq:Fdrainsph}--\ref{eq:Fsph} may also be
used if the galaxies are nearly spherical).
Before the hard binary stage, the BBH evolution timescales are obtained from
the equations in \S~\ref{sec:evlbfh} as in spherical galaxies.
We will first consider the evolution of BBHs with equal masses ($m_2/m_1=1$)
in the galaxies with different degrees of flattening $\epsilon$
(defined in \S~\ref{sec:Faxis}) (Fig.~\ref{fig:timeaxis});
and then consider the evolution of BBHs with different $m_2/m_1$ ratios for
a given value of $\epsilon(=0.3)$
(we estimated in \S~\ref{sec:Mlr} that the average $\epsilon$ in our sample
was $\sim0.36$) (Fig.~\ref{fig:timeaxis0.3}).

As seen from Figure~\ref{fig:mlrjrm} or \S~\ref{sec:Mlr}, if the BBH has equal
BH masses, the removed stellar mass from the galactic core during the non-hard
binary stage can be ignored and the loss wedge is approximately full when the
BBH becomes hard [$\eta(\E)\simeq1$ in equation \ref{eq:Mlw}].
As seen from Figure~\ref{fig:timeaxis} ($m_2/m_1=1$), the BBH evolution
timescales follow similar trends (small evolution timescales at large and
small radii, and bottlenecks at intermediate radii) to
those in spherical galaxies.
The bottleneck occurs during the hard binary stage.
After the BBHs become hard, the effects of flattening on the hardening
timescales depend on the value of $\epsilon$.
Figure~\ref{fig:timeaxis} shows that the bottleneck timescales in the
most nearly spherical galaxies
(e.g. $\epsilon$=0.01, 0.05 and 0.1 in panels a, b and c) are
still longer than the Hubble time,
but the bottleneck timescales in most highly flattened galaxies
(e.g. 21 out of our sample of 30 galaxies in panel d with $\epsilon=0.5$)
become shorter than the Hubble time.
If $\epsilon$ is small (e.g. $\la 0.01$), for most stars on
centrophilic orbits, the characteristic angular momenta $J\s$ (see
\S~\ref{sec:Faxis}) is not significantly larger than $J\lc(a)$ and the
evolution reduces to the spherical case (see most of the power-law galaxies or
some core galaxies in Fig.~\ref{fig:timeaxis}a and the galaxies in
Fig.~\ref{fig:timesph}a).
As $\epsilon$ increases, the effects of flattening become significant with
the increase of the total stellar mass in the loss wedge
(cf. Fig.~\ref{fig:mlrjrm}).
As seen from Figure~\ref{fig:timeaxis}(b)--(d),
the semimajor axes of the BBHs decrease more in axisymmetric galaxies than in
spherical galaxies before they reach the bottleneck stages,
and the bottleneck stages shift to smaller semimajor axes as $\epsilon$
increases.
For many BBHs in nearly spherical or low-$\epsilon$ axisymmetric galaxies,
(e.g. most of the BBHs in power-law galaxies in Fig.~\ref{fig:timeaxis}a--c or 
some BBHs in core galaxies in Fig.~\ref{fig:timeaxis}a whose evolution curves
have flat parts at intermediate radii $\sim 10^{-3}$--$1\pc$),
gravitational radiation becomes dominant still after the loss wedge/cone
is depleted; and the lifetime is controlled by the stellar refilling rate to
the loss wedge/cone.
As in spherical galaxies, after the loss wedge/cone is depleted, the inner
parts of most of the axisymmetric galaxies are also in the diffusion regime
before the gravitational radiation stage begins.
In the diffusion regime, the stellar diffusion rate to the loss wedge is also
insensitive to $a$, and hence the BBH evolution timescales controlled by the
stellar diffusion are nearly constant.
As seen from Figure~\ref{fig:timeaxis}(a)--(c), the flat parts of the evolution
curves in Fig.~\ref{fig:timeaxis} have almost the same heights as those in
Figure~\ref{fig:timesph}, and so the lifetimes of those BBHs are affected little
by the flattening effect, which is consistent with the results in MT that
the stellar tidal disruption rates by central BHs are not significantly
changed by the flattening effect.
With increasing $\epsilon$, more and more BBH evolution curves have no
flat parts, i.e., gravitational radiation becomes dominant before the loss
wedge is depleted (e.g. BBHs in power-law galaxies in Fig.~\ref{fig:timeaxis}d
and BBHs in core galaxies in Fig.~~\ref{fig:timeaxis}b-d);
and their merger timescales decrease, because the total stellar mass in the
loss wedge increases and gravitational radiation becomes dominant at smaller
semimajor axes.

Figure~\ref{fig:timeaxis0.3} shows the evolution for the BBHs with unequal BH
masses when the flattening parameter of the galaxies $\epsilon$ is 0.3.
In this case the removed stellar mass from the galactic core during the
non-hard binary stage can be ignored and the loss wedge is approximately full
when the BBHs (with $m_2/m_1\ga 0.01$) become hard
(see Fig.~\ref{fig:mlrjrm}e). 
As seen from Figure~\ref{fig:timeaxis0.3}, the bottleneck stages
(also in the hard binary stage) shift to smaller semimajor axes
as the mass ratio decreases, but not so obviously as those
in spherical cases (Fig.~\ref{fig:timesph}) since the stellar mass in the loss
wedges of the galaxies are generally so much not to be depleted
even when the BBHs reach the gravitational radiation stage.
As in spherical galaxies (Fig.~\ref{fig:timesph}),
the merger timescales decrease with decreasing mass ratios because
the gravitational radiation timescale $t_{\rm gr}$ (equation \ref{eq:tgr})
decreases as $a^4$.
When $m_2/m_1$ decreases to 0.01, most of the BBH merger timescales are shorter
than the Hubble time (Fig.~\ref{fig:timeaxis0.3}d). 

The justifications of our assumptions in \S~\ref{sec:testassmp} can be
generalized to axisymmetric galaxies (and triaxial cases below).
After the loss wedge is depleted, most of the stars contributing to
the BBH decay have energy $\E\peak\ll GM_\bullet/f_a a$ or
$r\peak\gg\max(r\coll,a\h)\ge f_a a$; the stellar diffusion timescales to the
loss wedges are smaller than the BBH hardening timescales;
and the removed stellar mass from the core is also less than the total core
mass.
If $\epsilon$ is small (e.g. $\epsilon=0.01$ in Fig.~\ref{fig:timeaxis}a),
the wandering of BBHs is not important, just as in spherical galaxies.
When $\epsilon$ is large (e.g. $\epsilon=0.3$ or $0.5$ in
Fig.~\ref{fig:timeaxis0.3} or \ref{fig:timeaxis}d),
the wandering of BBHs is also not important in core galaxies.
In a few of the power-law galaxies
whose lifetimes are longer than the Hubble time, $a=r_{\rm ran}$ happens
before the loss wedges are depleted and their BBH evolution timescales at
$a=r_{\rm ran}$ are shorter than the Hubble time, so the wandering of BBHs
might decrease the BBH lifetime in those galaxies. 

\subsection{Triaxial galaxies}\label{sec:restri}

\begin{figure}
\centerline{\psfig{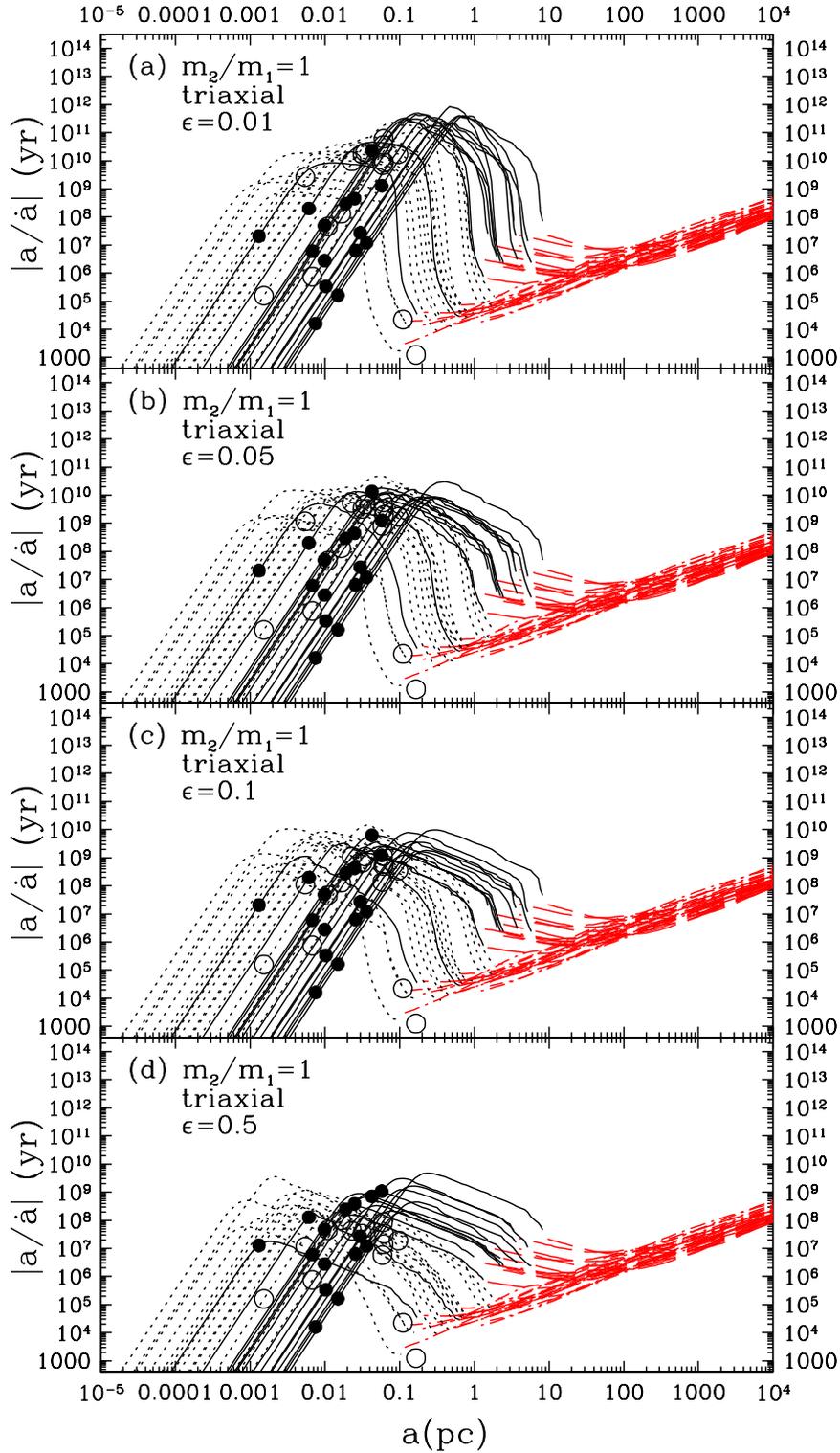}}
\caption{BBH evolution timescales for triaxial galaxies with $m_2/m_1=1$
and varying degrees of triaxiality $\epsilon$.
Secular evolution towards axisymmetry is not included.
The curves have the same meanings as those in Figure~\ref{fig:timesph}.
The solid circles (for core galaxies) and the open
circles (for power-law galaxies) represent the maximum Brownian motion
magnitudes of the BBH centers of mass.
Comparing with spherical galaxies and axisymmetric galaxies
(Fig.~\ref{fig:timesph} and \ref{fig:timeaxis}), the draining of the loss
regions in triaxial galaxies can significantly decrease the BBH lifetime.
Almost all the BBHs can merge within a Hubble time when $\epsilon$ is large
(panels c and d).
The lifetime of BBHs in the power-law triaxial galaxies with large $\epsilon$
(e.g. $\simgt 0.05$ in panels b--d) may be decreased by their Brownian motion.
}
\label{fig:timetri}
\end{figure}

\noindent
Using the equations in \S~\ref{sec:F} and \S~\ref{sec:evlbfh}, we obtain the
BBH evolution timescales if the galaxies in Table 1 are triaxial.
Here, we mainly consider the BBH with equal masses ($m_2/m_1=1$)
to see how the evolution depends on the triaxiality $\epsilon$.
The trends of the BBH evolution with different $m_2/m_1$ are
similar to those in axisymmetric galaxies (see Fig.~\ref{fig:timeaxis0.3}
or \S~\ref{sec:resaxis}).
Noting that triaxial galaxies containing central BHs are likely to evolve
secularly toward axisymmetry within a time $T^{\rm trans}$
(see \S~\ref{sec:Ftri}),
we will first ignore the evolution toward axisymmetry by setting
$T^{\rm trans}=\infty$ and see the maximum effect of the triaxiality on
the BBH evolution. The results are shown in Figure~\ref{fig:timetri}.
Then, we will set finite $T^{\rm trans}$ to see the effect of triaxiality
on the BBH evolution.
The age of galactic triaxiality $T\age^{\rm tri}$ is assumed to be
zero at $a=a\h$.

Figure~\ref{fig:timetri} shows the BBH evolution in triaxial galaxies without
including secular evolution towards axisymmetry.
As seen from Figure~\ref{fig:timetri}, the BBH evolution curves in triaxial
galaxies follow similar trends to those in spherical or axisymmetric galaxies
and the effects of triaxiality on the evolution timescales also depend on
the value of $\epsilon$ as in axisymmetric galaxies.
If $\epsilon$ is small, the characteristic angular momentum $J\s$ is not
significantly larger than $J\lc$ and the BBH evolution would be similar to that
in spherical or axisymmetric cases.
For example, when $\epsilon=0.01$ (Fig.~\ref{fig:timetri}a), most of the
BBH evolution timescales of the power-law galaxies are like those in the
spherical or axisymmetric galaxies and the bottleneck stages are
controlled by the stellar diffusion to the loss cones/wedges.
When $\epsilon$ is large ($\ga 0.05$, see Fig.~\ref{fig:timetri}b--d),
the difference between the loss region and the loss cone/wedge
becomes significant. The loss regions in all the galaxies cannot be
depleted before the gravitational radiation stage begins because there are
many more stars in the loss regions $J<J\s$ than in the loss cones/wedges.
Most of the BBH lifetimes in Figure~\ref{fig:timetri}(b) ($\epsilon=0.05$) are
comparable to or shorter than the Hubble time
and almost all the BBHs in  Figure~\ref{fig:timetri}(c)--(d)
($\epsilon=0.1$ and 0.5) can merge within a Hubble time.
As in axisymmetric galaxies, only the lifetime of BBHs in power-law triaxial
galaxies with large $\epsilon$ ($\epsilon\ga 0.05$ in panels b--d)
might be decreased by their wandering motion. 

After including secular evolution towards axisymmetry, the effect of
triaxiality on the BBH evolution depends on the value of $T^{\rm trans}$.
For example, if $T^{\rm trans}=10^2P(\E)$, after the BBHs become hard,
they first follow similar evolution to those in Figure~\ref{fig:timetri};
then with evolution towards axisymmetry, they evolve as those in axisymmetric
galaxies (Fig.~\ref{fig:timeaxis}); and the BBHs
have similar lifetimes to those in axisymmetric galaxies.
If $T^{\rm trans}$ is smaller, the BBH evolution curves would be close to
those in axisymmetric galaxies at earlier time; and if $T^{\rm trans}$ is
larger, they would be more like those in Figure~\ref{fig:timetri}.
Actually, if $T^{\rm trans}=10^3P(\E)$, the BBH evolution curves in
triaxial galaxies with large $\epsilon$ (e.g. $\ga 0.05$) are quite similar
to those in Figure~\ref{fig:timetri}(b)--(d).

In the discussion below on BBHs in triaxial galaxies, we will always use
the results in Figure~\ref{fig:timetri}, which gives the maximum effect of
triaxiality on BBH evolution.

\subsection{Generalization to a distribution of stellar masses and radii}\label{sec:gendis}

\noindent
In the calculations above, we assume a single stellar mass of $1\msun$ and a
single stellar radius of
$1\rsun$.  In this subsection, we will see that generalizing to a distribution
of stellar masses and radii does not significantly affect our results.

First, generalizing to a range of stellar mass does not affect the dynamical
friction timescales (see equations \ref{eq:fricforce}--\ref{eq:frictime}),
which is affected by the mass density, not the mass of a single star.
The value of $\Lambda$ may be sensitive to the stellar mass, but the friction
timescale depends only logarithmically on $\Lambda$.

Second, in the hard binary stage, the BBH hardening timescale is controlled by
the mass clearing rate from the loss region (see equation \ref{eq:th}).  If
the loss region is not depleted, generalizing to a range of stellar mass does
not affect the BBH hardening timescale since the total mass cleared from the
loss region per unit time is not changed.  Only when the evolution is
controlled by stellar diffusion to the loss region is the BBH evolution
timescale affected through the variation of two-body relaxation diffusion
coefficient with stellar mass (see equation \ref{eq:mu}).
According to equation (A6) in MT, the diffusion coefficient of two-body
relaxation at constant mass density $\oa\mu\propto
\int_{m_1}^{m_2}m_*^2n(m_*)\d m_*$, where $n(m_*)$ is proportional to the
probability of finding a star with mass $m_*\rightarrow m_*+\d m_*$ and
$\int_{m_1}^{m_2}n(m_*)\d m_*=\msun$.
Thus, if we use a Salpeter mass function $n(m_*)\propto m_*^{-2.35}$ with
limits $m_1=0.08\msun$ and $m_2=1\msun$, the diffusion coefficient
is reduced by a factor of 0.31 compared to a population of stars of mass
$1\msun$.
Therefore, when the stellar diffusion to the loss cone or loss wedge controls
the BBH evolution timescales, the evolution timescales will be only
increased by a factor of three or so, which does not affect our qualitative
conclusions.

Third, if the loss region is depleted in the non-hard binary stage, the
evolution of the BBH after the depletion and before the hard binary stage is
controlled by both dynamical friction from distant stars and three-body
interactions with low-angular momentum stars diffused from high-angular
momentum orbits, which further reduces the effects of the dependence of the
diffusion coefficient on the stellar mass distribution.

Finally, the range of stellar radii has relatively little effect on the
collision radius radius $r_{\rm coll}$ (equation \ref{eq:rcoll}).
Thus, we still have $r\coll\ll r\peak$ in Figure~\ref{fig:Mr} and the BBH
evolution timescale is not affected much (see \S~\ref{sec:Mlr}).

Since generalizing to a distribution of stellar masses and radii does not
significantly affect the BBH evolution timescales, we will continue to
use the results with a single stellar mass and radius (obtained in
\S~\ref{sec:ressph}--\S~\ref{sec:restri}) to study the properties of surviving
BBHs in \S~\ref{sec:bbhpro}.

\section{Properties of surviving BBHs}\label{sec:bbhpro}

\noindent
As seen from \S~\ref{sec:res}, BBHs can survive for a Hubble time in spherical
galaxies, axisymmetric galaxies or some galaxies with small triaxiality
($\epsilon\la 0.05$).  Whether or not some of the various astronomical
phenomena that have been associated with BBHs [e.g. double nuclei;
double-peaked Balmer lines \cite{G96}; quasi-periodic radio, optical, X-ray
or $\gamma$-ray variability \cite{Setal88,RM00} etc.]
can be observed depends on the current
orbital properties of the BBHs, such as their semimajor axis $a$, circular
speed $v\c=\sqrt{GM_\bullet/a}$, orbital period $P_{\rm orb}=2\pi a/v\c$, etc.

\begin{figure}
\centerline{\psfig{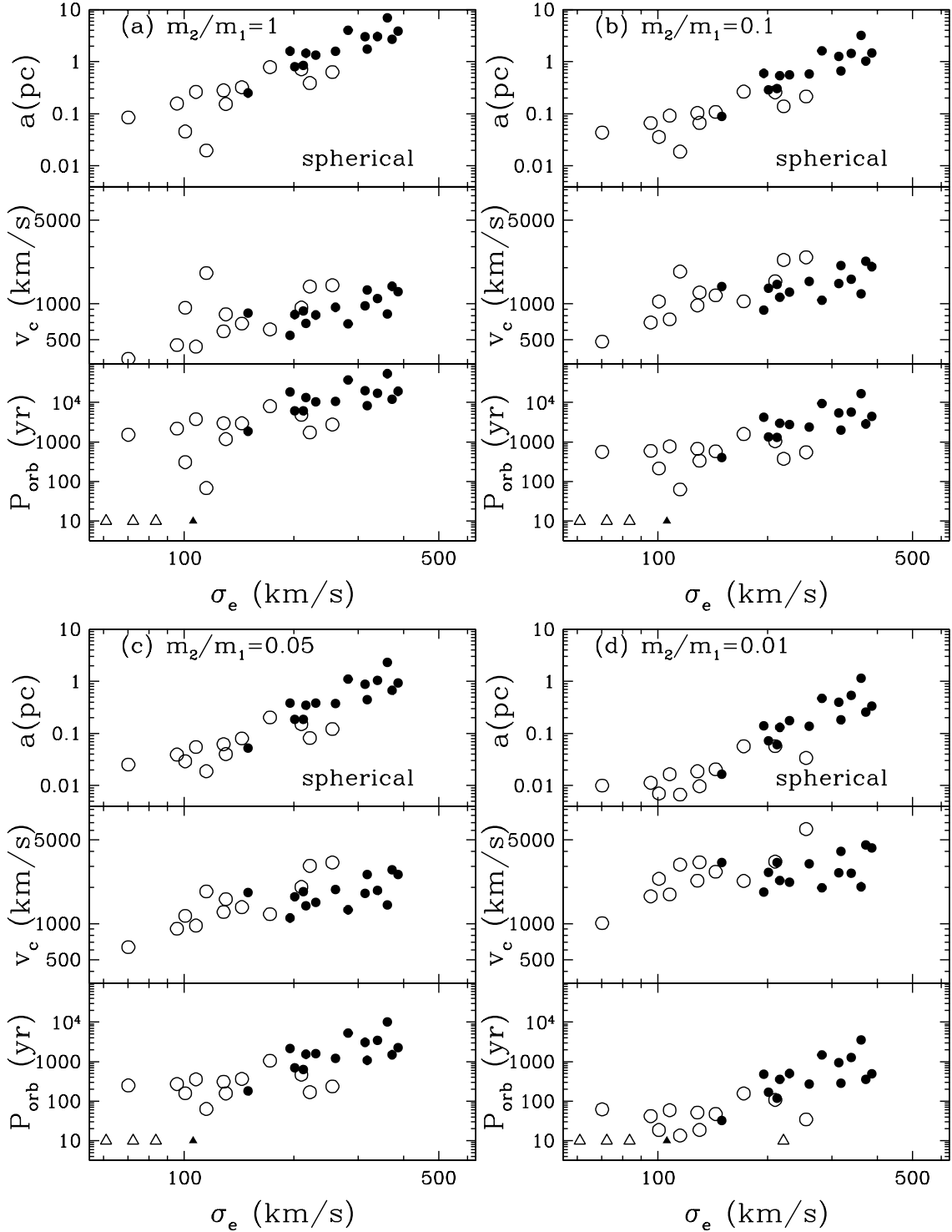}}
\caption{Estimated observational properties of surviving BBHs in spherical
galaxies with varying mass ratios.
The panels show the relations between the galactic velocity
dispersion $\sigma\e$ and the BBH semimajor axes $a$, circular speeds
$v\c=\sqrt{GM_\bullet/a}$ and orbital periods $P_{\rm orb}=2\pi a/v\c$
at the point where $|a/\dot a|=10^{10}$ years for the first time in
Figure~\ref{fig:timesph}.
The solid circles represent the core galaxies and the
open circles represent the power-law galaxies.
The solid triangles (core galaxies) and open triangles (power-law galaxies)
located at the bottom of each panel represent the dispersion of the galaxies
in which the BBH evolution timescales are less than the Hubble time
$10^{10}$ years.
BBHs in most of the low-dispersion ($\sigma\e\simlt 90\kms$) galaxies have
merged within a Hubble time and BBHs are more likely to survive in the
galaxies with $\sigma\e\simgt 90\kms$
(similarly in Fig.~\ref{fig:avtsigmaaxis}--\ref{fig:avtsigmatri} below).
The semimajor axes and orbital periods of surviving BBHs are larger in
high-dispersion galaxies than in low-dispersion galaxies (similarly in
Fig.~\ref{fig:avtsigmaaxis}--\ref{fig:avtsigmatri} below) and larger for
BBHs with equal masses than for BBHs with unequal masses (similarly in
Fig.~\ref{fig:avtsigmaaxis0.3} below).
The circular speeds are larger in high-dispersion galaxies than in
low-dispersion galaxies and larger for BBHs with unequal masses than
for BBHs with equal masses (similarly in Fig.~\ref{fig:avtsigmaaxis0.3} below).
Note that BBHs with very low mass ratios (say, $\simlt 0.001$) arising from
mergers of galaxies would not exist at galactic centers because the smaller
BH cannot sink into the galactic center within a Hubble time
(cf. \S~\ref{sec:resspheqms}).
}
\label{fig:avtsigmasph}
\end{figure}

Figure~\ref{fig:avtsigmasph} shows the relation between the galactic velocity
dispersions $\sigma\e$ and the BBH orbital parameters at the point where
$|a/\dot a|=10^{10}\yr$ as $a$ decreases for the first time in
Figure~\ref{fig:timesph} (spherical galaxies).
As seen from Figure~\ref{fig:avtsigmasph}, the BBH orbital parameters depend
on galactic velocity dispersions and BH mass ratios.
The BBHs in most of the galaxies with $\sigma\e\la 90\kms$ have merged.
The semimajor axes and orbital periods of surviving BBHs are generally in the
range $10^{-3}$--$10\pc$ and 10--$10^5\yr$; and they are generally larger in
high-dispersion galaxies (core galaxies with high central BH masses) than
in low-dispersion galaxies (power-law galaxies with low central BH masses),
and larger for BBHs with equal BH masses than for BBHs with unequal masses.
The orbital velocities of surviving BBHs are generally in the range
$10^2$--$10^4\kms$; and they are generally larger in high-dispersion galaxies
than in low-dispersion galaxies and larger for BBHs with unequal masses
than for BBHs with equal masses.
The surviving BBH properties estimated above may help identify appropriate
methods to probe BBHs in different spherical galaxies:
\begin{enumerate}

\item Double nuclei associated with BBHs should be easier to observe in
high-dispersion or luminous galaxies than in low-dispersion or faint galaxies
because BBH semimajor axes are much larger in high-dispersion galaxies.
The BBH sample identified from double nuclei would
have a bias toward high-$m_2/m_1$ (say, $\ga 0.1$) BBHs for at least three
reasons: (1) the two components have comparable brightness if the stellar
or non-stellar luminosity is correlated with the BH mass, (2) the BBH
semimajor axes are generally larger if the BH masses are equal.  (3) BBHs with
very low-$m_2/m_1$ ratios (say, $\la 0.001$) may not exist in the galactic
center since the smaller BH may not sink to the galactic center in a Hubble
time (see Fig.~\ref{fig:timesph}d) (here
we only consider the BBH formed by mergers of galaxies, and we do not consider
the possible low-mass BH formation caused by some mechanism at galactic
centers).

\item Broad line regions (BLRs) associated with the two components of a
BBH in AGNs may produce double-peaked emission lines (e.g. Balmer lines)
varying with timescales of the BBH orbital period, though the double peaks are
not easily recognized for $v\c\la\sigma_{\rm BLR}$ where $\sigma_{\rm BLR}$ is
the velocity dispersion of the clouds in BLRs (typically 3000--5000 $\kms$).
Though the BBH orbital period can be as short as 10 yr for $m_2/m_1=0.01$
(Fig.~\ref{fig:avtsigmasph}d), for such an extreme mass ratio one component of
double peaks would be too weak to be discernible compared to the other strong
component.  The double peaks from
the BLRs of high-$m_2/m_1$ BBRs would have comparable strengths, but it is
hard to detect the line variability within a reasonable
time for high-$m_2/m_1$ BBHs since $P_{\rm orb}$ is at least
$10^2$--$10^3\yr$ (see the case $m_2/m_1=0.1$ in Fig.~\ref{fig:avtsigmasph}b)
even in low-dispersion galaxies ($\sigma\e\la150\kms$).
So, double-peaked emission lines from BLRs are unlikely to show periodic
behavior due to the long orbital period in BBH galaxies.
For example, the quasar 3C390.3 was introduced as a BBH candidate based on
data from only $\sim20$ years, which is much less than the estimated BBH
orbital period $P_{\rm orb}\sim 300\yr$.
The radial velocity curve of $\sim 20\yr$ for one peak of emission line
H$\beta$ in 3C390.3 is consistent with the
expectation from the orbital motion of a BBH \cite{G96};
but the results from continuous observations of $\sim 10$ more years deviate
from the expectation from a BBH model \cite{EHG97}.

\item Some periodic behavior observed in the radio,
optical, X-ray or $\gamma$-ray lightcurves is possibly related to a BBH
creating jet(s) aligned nearly along the line of sight or interacting with
disk(s).  The observed period is usually identified with the orbital period of
the BBH or a fraction of the orbital period.  For example, if a relativistic
jet emerges from the less massive BH and is aligned nearly along the line of
sight, the flux of the X/$\gamma$-ray radiation
emitted from the jet may vary with the periodic change of the
Doppler-boosting factor (due to the slight change of the jet inclination angle
with the BH orbital motion) \cite{RM00}.  Because of the same mechanism in
the phenomenon of ``superluminal motion'' (i.e. the observed timescale is
shorter than the intrinsic timescale, see Rees 1966), the observed
X/$\gamma$-ray variability timescale is shorter than the orbital period
[say, $\sim 10^{-2}P_{\rm orb}$ in Rieger \& Mannheim (2000)] which can
be identified within a much shorter time ($\sim$ 0.1--1$\yr$) in low-dispersion
galaxies with $m_2/m_1\sim 0.01$ ($P_{\rm orb}\sim10$--$100\yr$ when
$\sigma\e\la150\kms$ in Fig.~\ref{fig:avtsigmasph}d).  It is hard to use other
variability phenomena with timescales $\sim P_{\rm orb}$ to search for
BBHs within a short time (say, $<50\yr$) except for low-$m_2/m_1$ ($\sim 0.01$)
ones in low-dispersion galaxies ($\sigma\e\la150\kms$).

\end{enumerate}

\begin{figure}
\centerline{\psfig{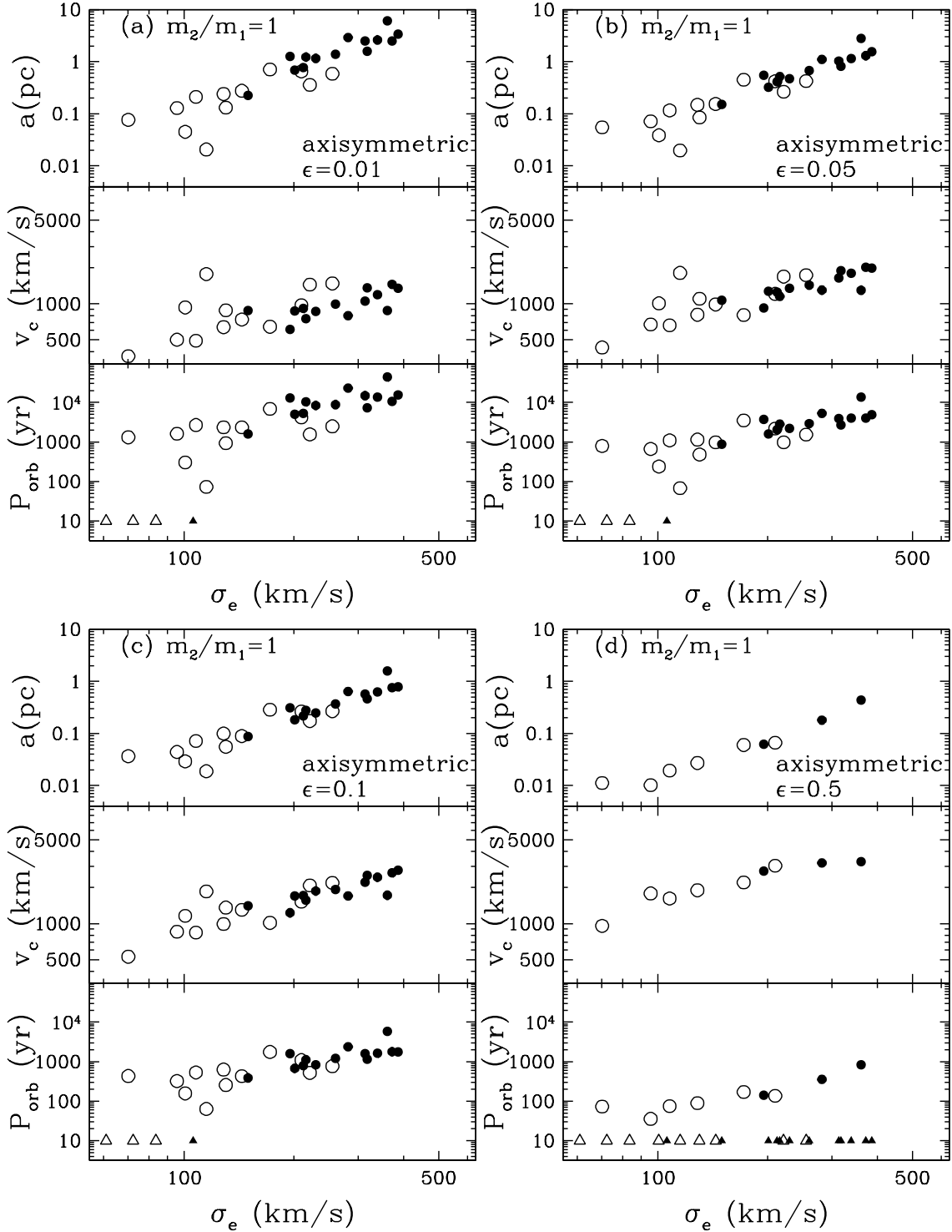}}
\caption{Estimated observational properties of surviving BBHs in axisymmetric
galaxies with $m_2/m_1=1$ and varying degrees of flattening $\epsilon$
(see Fig.~\ref{fig:timeaxis}). The symbols have the same meanings as those in
Fig.~\ref{fig:avtsigmasph}.
The relations between the surviving BBH orbital properties and the galactic
dispersion follow similar trends to the relations in spherical galaxies
(Fig.~\ref{fig:avtsigmasph}).
Most BBHs in highly flattened galaxies can merge within a Hubble time (panel d).
The semimajor axes and the orbital periods of surviving BBHs are larger
in nearly spherical galaxies than in highly flattened galaxies;
while the circular speeds are larger in highly flattened galaxies than
in nearly spherical galaxies (similarly in Fig.~\ref{fig:avtsigmatri} below).
In power-law or low-dispersion galaxies with large $\epsilon$
(e.g. panel c or d), Brownian motion of BBHs may further
decrease the semimajor axis (at the same time, decrease the orbital period and
increase the circular speed) shown in the figure or make the BBHs merge within
a Hubble time
(similarly in Fig.~\ref{fig:avtsigmaaxis0.3}--\ref{fig:asigma} below).
}
\label{fig:avtsigmaaxis}
\end{figure}

\begin{figure}
\centerline{\psfig{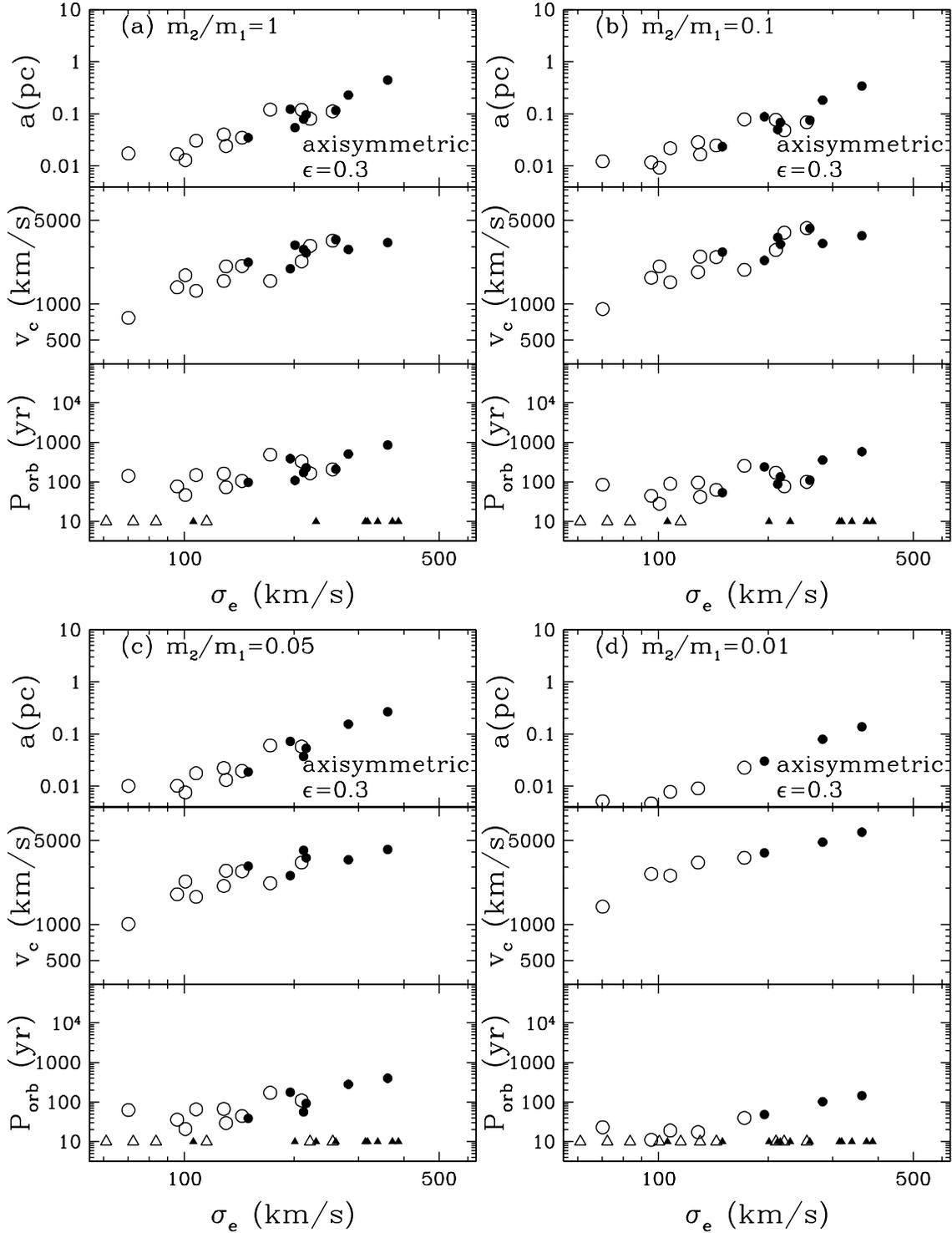}}
\caption{Estimated observational properties of surviving BBHs in axisymmetric
galaxies with $\epsilon=0.3$ and varying mass ratios (see
Fig.~\ref{fig:timeaxis0.3}). The symbols have the same meanings as those in
Fig.~\ref{fig:avtsigmasph}.
The relations between the surviving BBH orbital properties and mass ratios
follow similar trends to those in spherical galaxies
(Fig.~\ref{fig:avtsigmasph}).
With decreasing BH mass ratios, more and more BBHs can merge within a Hubble
time.
}
\label{fig:avtsigmaaxis0.3}
\end{figure}

\begin{figure}
\centerline{\psfig{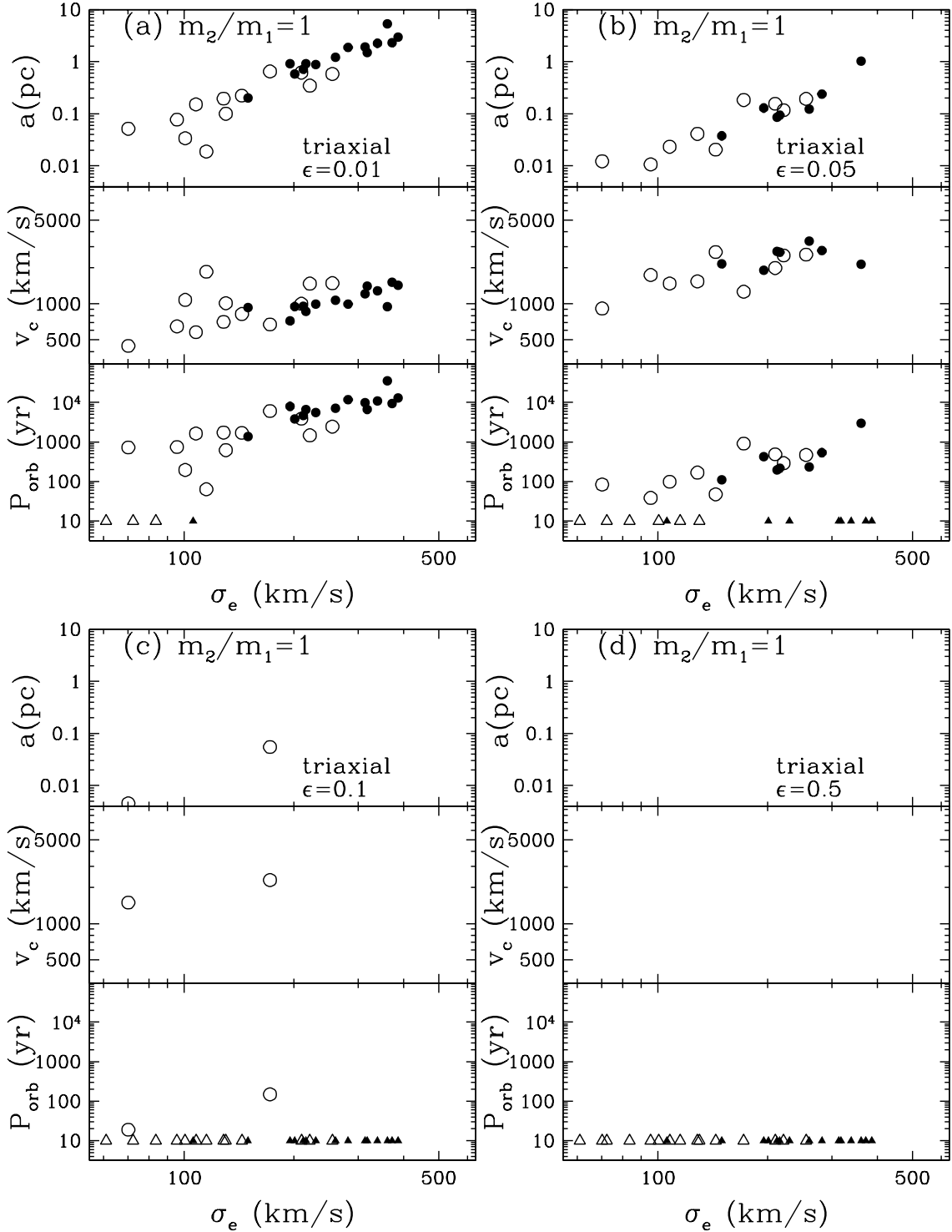}}
\caption{Estimated observational properties of surviving BBHs in triaxial
galaxies with $m_2/m_1=1$ and varying degrees of triaxiality $\epsilon$
(see Fig.~\ref{fig:timetri}) and without including secular evolution
towards axisymmetry.
The symbols have the same meanings as those in Figure~\ref{fig:avtsigmasph}.
Except for some BBHs in weakly triaxial galaxies
(e.g. $\epsilon=0.01$ or 0.05 in panel a or b), most BBHs in triaxial
galaxies can merge within a Hubble time.
}
\label{fig:avtsigmatri}
\end{figure}

In axisymmetric or weakly triaxial ($\epsilon\la 0.05$) galaxies
(Fig.~\ref{fig:avtsigmaaxis}--\ref{fig:avtsigmatri}),
the relations between the orbital parameters of surviving BBHs and the
properties of their host galaxy follow similar trends to the relations in
spherical galaxies (Fig.~\ref{fig:avtsigmasph}), though the normalization
may be different.
The orbital properties of surviving BBHs are in the similar range as those
in spherical galaxies, and they depend on the flattening or triaxiality
parameter $\epsilon$, as well as the galactic velocity dispersion and BH
mass ratio.
As seen from Figure~\ref{fig:avtsigmaaxis} (BBHs with mass ratio
$m_2/m_1=1$ in axisymmetric galaxies), with increasing flattening parameter
$\epsilon$, BBH semimajor axes decrease (the upper limit is $\sim 10\pc$ for
$\epsilon=0.01$
and $\sim 1\pc$ for $\epsilon=0.5$), which may require higher resolution to
resolve double nuclei;
BBH orbital periods decrease ($\sim 10^2$--$10^5\yr$ for $\epsilon=0.01$ and
$\sim 50$--$10^3\yr$ for $\epsilon=0.5$),
which may help in the identification of 
periodicities in X/$\gamma$-ray variability within a short time
($\sim 0.5$--$10\yr$ if the timescale of periodicities
$\sim 10^{-2} P_{\rm orb}$), but
the orbital periods are still so long that the periodic variability of
double-peaked emission lines from BLRs will be difficult to identify. 
As seen from Figure~\ref{fig:avtsigmaaxis0.3}
(flattened galaxies with $\epsilon=0.3$),
with decreasing $m_2/m_1$ ratios, the BBH semimajor axes and the orbital
periods decrease as in spherical galaxies.  When $m_2/m_1$ decreases to 0.01
(Fig.~\ref{fig:avtsigmaaxis0.3}d),
many BBHs (22 out of our sample of 30 galaxies) have merged within a Hubble
time; and most of the surviving BBHs in low-dispersion galaxies
($\sigma\e\la 150\kms$) have an orbital period $\sim$10--30$\yr$ and could
possibly be found by
identifying some variability phenomenon associated with the orbital motion.
It is difficult to observe BBHs even with equal masses in more triaxial
galaxies ($\epsilon\ga 0.1$) because most of them have merged within a Hubble
time (Fig.~\ref{fig:avtsigmatri}c and d). 
In addition, the random motion of the center of mass of BBHs might affect
their observational properties in low-dispersion axisymmetric galaxies (with
$\epsilon\ga 0.1$) or low-dispersion triaxial galaxies
(with $\epsilon\sim 0.05$).
On one hand, the random motion may decrease their semimajor axes and orbital
periods,
which would make variability phenomena a more efficient tool to search
for BBH candidates;
on the other hand, it may also make more BBHs merge within a Hubble time
and decrease the number of surviving BBHs, which would
decrease the probability of identifying BBHs by variability phenomena.

Note that the double-peaked emission lines from BLRs and radio, optical,
X/$\gamma$-ray variability that we have discussed above are phenomena seen
in the relatively rare and distant AGNs, not the common inactive nearby
galaxies where BHs are known to exist.

If future observations do not reveal any phenomena associated with BBHs
in AGNs, that might suggest that: (i) the inner stellar distribution in AGNs
is different from normal nearby early-type galaxies,
(ii) gas in AGNs plays
an important role in BBH orbital decay (e.g. Gould \& Rix 2000),
or (iii) nuclear activity switches on
soon after a galaxy merger and the lifetime of nuclear activity is
much less than the Hubble time, so the surviving BBHs in AGNs would
not have had time to spiral together before the nuclear activity
ceases.

\begin{figure}
\centerline{\psfig{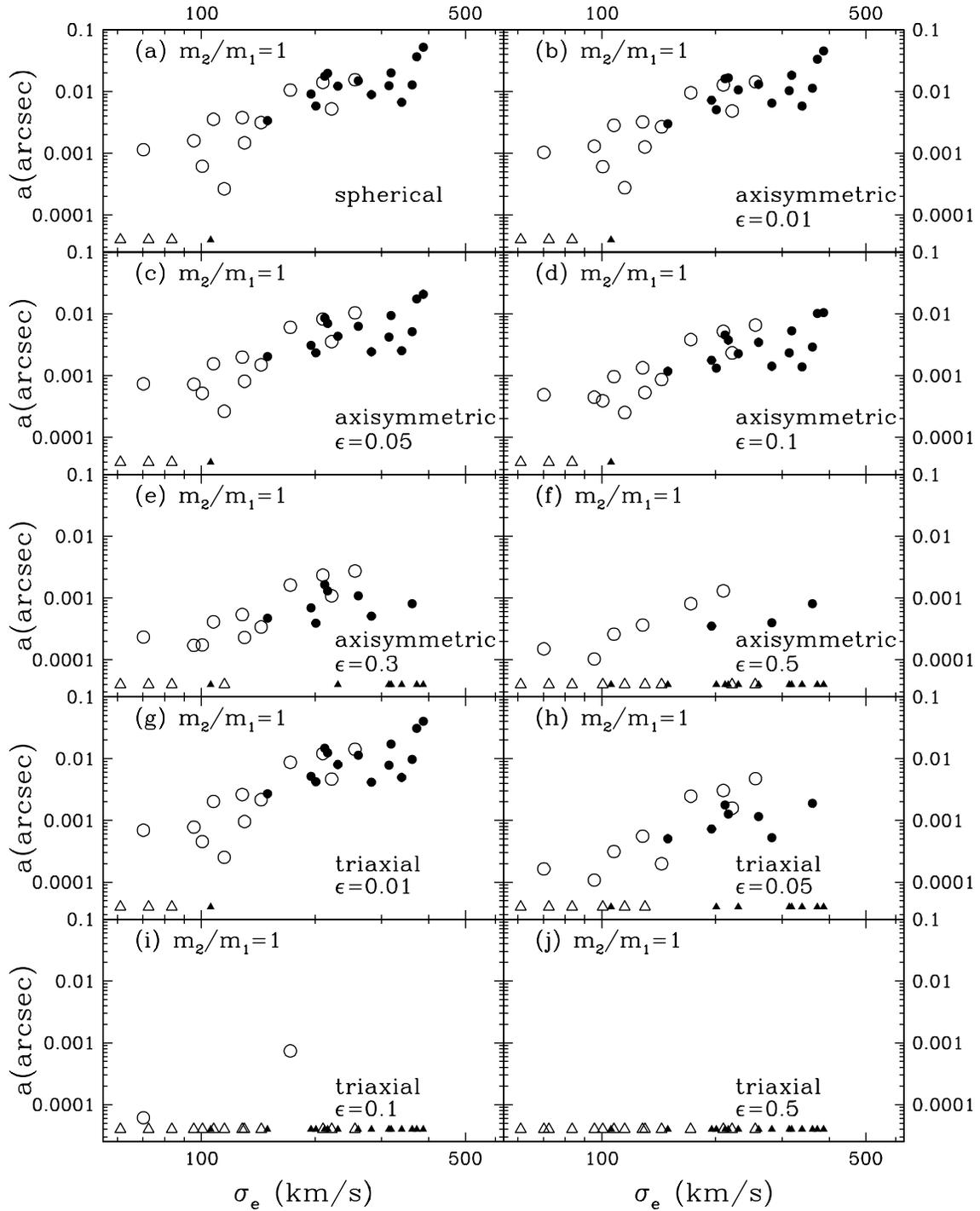}}
\caption{Estimated semimajor axes {\it in arcsec} of surviving BBHs
(with $m_2/m_1=1$) versus galactic velocity dispersion $\sigma\e$.
The symbols have the same meanings as in Figure~\ref{fig:avtsigmasph}.
The galaxies are assumed to be spherical in panel (a), axisymmetric
in panels (b)--(f) and triaxial (without including secular evolution towards
axisymmetry) in panels (g)--(j).
The upper limit of the BBH semimajor axes is just within the {\it HST}
resolution, 0.1\arcsec.
}
\label{fig:asigma}
\end{figure}

To see if double nuclei associated with BBHs formed by mergers of galaxies
can be observed with current
telescope resolution, we show the predicted BBH semimajor axes {\it in arcsec}
versus the galactic velocity dispersion $\sigma\e$ for the nearby inactive
galaxies (Table 1) in Figure~\ref{fig:asigma}.
The BBHs are assumed to have equal BH masses, which have larger semimajor
axes than those with unequal masses.
As seen from Figure~\ref{fig:asigma}, spherical or nearly spherical and
high-dispersion galaxies have surviving BBHs with the largest semimajor axes, 
which are just within the {\it HST} resolution, 0.1\arcsec.

\section{Discussion and Conclusions}\label{sec:discon}

\noindent
We have studied the evolution of BBHs in realistic galaxy models
obtained from a sample of nearby early-type galaxies.  We calculated the BBH
evolution timescales due to stellar interactions, taking into account the
refilling of the loss cone by two-body
relaxation and by tidal forces in non-spherical galaxies.  The
evolution of BBHs depends on BH mass ratio and host galaxy type.  BBHs with
low mass ratios (say, $m_2/m_1\la0.001$) are
only rarely formed by mergers of galaxies because
the dynamical friction timescale is too long for the
smaller BH to sink into the galactic center within a Hubble time
(cf. Fig.~\ref{fig:timefric}).  In spherical galaxies, if the BBH has equal BH
masses, low-angular momentum stars are generally depleted quickly after the
BBH becomes hard and thereafter the BBH lifetime is controlled by the
stellar diffusion rates into the loss cone (see
Figs.~\ref{fig:mlrjrm}a, \ref{fig:timesph}a and \ref{fig:rlca}); if the BBH
has unequal masses (so that the BBH becomes hard at smaller
semimajor axis), low-angular momentum stars can be depleted before the BBH
becomes hard, and possibly gravitational radiation can dominate the BBH
orbital decay in the non-hard binary stage and the BBH lifetime decreases
(see Figs.~\ref{fig:mlrjrm}a and \ref{fig:timesph}b--d).
Low-dispersion galaxies have smaller central BHs and higher central densities,
and hence their BBHs generally have shorter hardening timescales
(associated with interactions with stars) in the hard binary stage
(see equation \ref{eq:th}) and shorter lifetimes (see Fig.~\ref{fig:timesph}).
In axisymmetric and triaxial galaxies,
many stars can precess onto the low-angular momentum orbits by tidal forces
and decrease the lifetime of BBHs (see Figs.~\ref{fig:mlrjrm}b--f,
\ref{fig:timeaxis}--\ref{fig:timetri}).  The wandering of the BBHs is
generally not important, but might decrease the lifetime of BBHs in some
axisymmetric or triaxial galaxies with low velocity dispersion (see
Figs.~\ref{fig:timeaxis}--\ref{fig:timetri}).
Note that in the study, stars escaping from the core are considered as removed
from the loss cone, which is usually a plausible assumption;
but in some circumstances, even a small fraction of stars returning to the
loss cone could enhance the BBH decay rate, which deserves further
investigation.

Most of BBHs in the galaxies with velocity dispersion $\sigma\e\la90\kms$
have merger timescales shorter than a Hubble time
(see Fig.~\ref{fig:avtsigmaaxis}--\ref{fig:avtsigmatri}), and so do
BBHs in highly flattened ($\epsilon\ga 0.5$) or moderately triaxial
($\epsilon\ga0.05$) galaxies
(see Figs.~\ref{fig:avtsigmaaxis}d and \ref{fig:avtsigmatri}b--d).
BBHs with low-$m_2/m_1$ (say, $\sim 0.01$) ratios, which have shorter lifetime
than the BBHs with equal masses, can also merge within a Hubble time in
axisymmetric or triaxial galaxies (see Fig.~\ref{fig:avtsigmaaxis0.3}d).
Surviving BBHs are generally in the galaxies with velocity dispersion
$\sigma\e\ga 90\kms$.
Spherical galaxies, axisymmetric or weakly triaxial galaxies are all
likely to have surviving BBHs, especially have those with equal masses
(see Figs.~\ref{fig:avtsigmasph}--\ref{fig:avtsigmaaxis0.3} and
\ref{fig:avtsigmatri}a--b).
We also estimated the observational properties of surviving BBHs.
The semimajor axes and orbital periods of surviving BBHs are generally in the
range $10^{-3}$--$10\pc$ and 10--$10^5\yr$;
and they are generally larger in high-dispersion galaxies than in
low-dispersion galaxies, larger in spherical galaxies than in non-spherical
galaxies, and larger for BBHs with equal BH masses than for BBHs with
unequal masses.
The orbital velocities of surviving BBHs are generally in the range
$10^2$--$10^4\kms$; and they are generally larger in high-dispersion galaxies
than in low-dispersion galaxies, larger in non-spherical galaxies than in
spherical/nearly spherical galaxies and larger for BBHs with unequal masses
than for BBHs with equal masses.

In short, our study shows that BBHs arising from galaxy mergers are likely to
have merged in low-dispersion or highly flattened/triaxial galaxies, but
should still survive in spherical/nearly spherical and high-dispersion
galaxies. The study in this paper may help to further explore the merger
history of massive BHs.

The BBH mergers driven by stellar dynamics are also simply discussed by
Gould \& Rix (2000).
They conclude that mergers driven by stellar dynamics are nearly impossible in
a Hubble time and present a mechanism by which gas can drive BH mergers.
However, their conclusion is based on much simpler galaxy models and dynamics
than those used in this paper.

We have also discussed methods to detect surviving BBHs.
For inactive galaxies, currently, the only method to probe BBHs is to search
for double nuclei at galactic centers.
It would be easier to find double nuclei associated with BBHs in
luminous (or high-dispersion) nearly spherical galaxies and the BBHs to be
found would have a bias toward equal BH masses.
For active galaxies, we can also use periodic variability phenomena
to search for BBHs.
Within a short time (e.g. 1 month--1 yr), it would be easier to find 
X/$\gamma$-ray periodic variability associated with the BBH orbital motion
in faint or low-dispersion (e.g. $\sigma\e\la 150\kms$, but $\ga 90\kms$
since most BBHs have merged in the galaxies with $\sigma\e\la 90\kms$)
galaxies than in luminous or high-dispersion (e.g. $\sigma\e\ga 150\kms$)
galaxies.
Here, the variability timescale is usually a fraction of the BBH orbital period
(e.g. 0.01 times of the BBH orbital period),
and the BBHs to be found would have a bias toward low $m_2/m_1$
(e.g. $\sim0.01$) ratios.
By other variability phenomena with timescales of the BBH orbital
period, it is also easier to find the low-$m_2/m_1$ BBHs in faint or
low-dispersion ($\sigma\e\la 150\kms$, but $\ga 90\kms$) galaxies,
but it would take at least $\sim$10--100$\yr$.
It is hard to find BBHs by double-peaked emission lines
from BLRs either because it takes a long time ($\ga 50\yr$) to identify the
periodic variability of double peaks with timescales of BBH orbital periods or
because the strength of one component is too weak to separate it from the other
one.

The upper limit of the semimajor axes of surviving BBHs for all of the galaxies
in our study is $\sim 10\pc$, which is just within the {\it HST} resolution,
0.1\arcsec, or several pc for typical galaxies in the sample used in this
paper.
None of these nearby common inactive galaxies has shown signs of surviving
BBHs (close double nuclei with separation $0.1\arcsec$) at their centers.
It is not yet clear whether the absence of detected BBHs in nearby galaxies
presents a problem, and if so whether the problem is with the calculations in
this paper, the hierarchical structure formation model, or the observational
detection techniques.

\section*{Acknowledgments}
I am deeply indebted to my thesis advisor, Scott Tremaine, for his
encouragement and guidance.  The numerous suggestions and clarifications from
him have shaped up this work in the present form. I am also
very grateful to Jeremy Goodman and David N. Spergel for stimulating
discussions and suggestions, to Neta Bahcall for her care and encouragement in
the work.  This research was supported in part by NSF grant AST-9900316 and
NASA grant NAG5-7066.

\newpage
\begin{table*}
\begin{minipage}{165mm}
\begin{tabular}{lccccccccccccc}\hline \hline
Name & Type & $\log_{10}(r\b)$ & $\mu\b$ & $\alpha$ & $\beta$ & $\gamma$ &
$\log_{10}(r_{\rm e})$ & $\Upsilon_V$ & Distance & $M_\bullet$ & $\sigma\e$ \\
  &  & (pc) & & & & & (pc) & ($\msun/\Lsun$) & (Mpc) & ($\msun$) & $\kms$ \\
\hline
NGC596  & 1 & 2.56 & 18.03 & 0.76 & 1.97 & 0.55 & 3.49 &  4.2 & 21.2 & 3.5e7 & 140 \\
NGC720  & 1 & 2.55 & 17.50 & 2.32 & 1.66 & 0.06 & 3.64 &  8.2 & 22.6 & 2.0e8 & 230 \\
NGC1172 & 1 & 2.55 & 18.61 & 1.52 & 1.64 & 1.01 & 3.75 &  2.6 & 29.8 & 4.5e6 & 84 \\
NGC1399 & 1 & 2.43 & 17.06 & 1.50 & 1.68 & 0.07 & 3.56 & 12.7 & 17.9 & 6.8e8 & 320 \\
NGC1426 & 1 & 2.23 & 17.53 & 3.62 & 1.35 & 0.85 & 3.44 &  4.9 & 21.5 & 2.4e7 & 130  \\
NGC1600 & 1 & 2.88 & 18.38 & 1.98 & 1.50 & 0.08 & 4.06 & 14.3  & 50.2 & 6.5e8 & 310 \\
NGC3115 & 2 & 2.07 & 16.17 & 1.47 & 1.43 & 0.78 & 3.17 &  7.1 &  8.4 & 3.0e8 & 260  \\
NGC3377 & 1 & 0.64 & 12.85 & 1.92 & 1.33 & 0.29 & 3.21 &  2.9 &  9.9 & 1.1e7 & 110  \\
NGC3379 & 1 & 1.92 & 16.10 & 1.59 & 1.43 & 0.18 & 3.23 &  6.9 &  9.9 & 1.5e8 & 210  \\
NGC3599 & 2 & 2.12 & 17.58 &13.01 & 1.66 & 0.79 & 3.47 &  2.1 & 20.3 & 1.4e6 & 61  \\
NGC3605 & 1 & 1.94 & 17.25 & 9.14 & 1.26 & 0.67 & 3.23 &  4.1 & 20.3 & 7.5e6 & 96  \\
NGC4168 & 1 & 2.65 & 18.33 & 0.95 & 1.50 & 0.14 & 3.90 &  7.5 & 36.4 & 1.1e8 & 200  \\
NGC4239 & 4 & 1.98 & 18.37 &14.53 & 0.96 & 0.65 & 3.08 & 3.4  & 15.3 & 2.4e6 & 70   \\
NGC4365 & 1 & 2.25 & 16.77 & 2.06 & 1.27 & 0.15 & 3.79 &  8.4 & 22.0 & 3.2e8 & 260  \\
NGC4387 & 1 & 2.52 & 18.89 & 3.36 & 1.59 & 0.72 & 3.06 &  5.3 & 15.3 & 1.2e7 & 110  \\
NGC4434 & 1 & 2.25 & 18.21 & 0.98 & 1.78 & 0.70 & 3.14 &  4.7 & 15.3 & 9.1e6 & 100  \\
NGC4458 & 1 & 0.95 & 14.49 & 5.26 & 1.43 & 0.49 & 3.30 &  4.0 & 15.3 & 2.7e6 & 72  \\
NGC4464 & 1 & 1.95 & 17.35 & 1.64 & 1.68 & 0.88 & 2.60 &  4.8 & 15.3 & 1.5e7 & 120  \\
NGC4478 & 1 & 1.10 & 15.40 & 3.32 & 0.84 & 0.43 & 3.02 &  5.0 & 15.3 & 6.8e7 &
170  \\
NGC4486 & 1 & 2.75 & 17.86 & 2.82 & 1.39 & 0.25 & 3.89 & 17.7  & 15.3 & 1.4e9 & 390  \\
NGC4551 & 1 & 2.46 & 18.83 & 2.94 & 1.23 & 0.80 & 3.12 &  7.3 & 15.3 & 2.3e7 & 130  \\
NGC4564 & 1 & 1.59 & 15.70 & 0.25 & 1.90 & 0.05 & 3.21 & 4.8  & 15.3 & 4.0e7 & 150  \\
NGC4621 & 1 & 2.34 & 17.20 & 0.19 & 1.71 & 0.50 & 3.54 &  6.7 & 15.3 & 1.8e8 & 220  \\
NGC4636 & 1 & 2.38 & 17.72 & 1.64 & 1.33 & 0.13 & 3.88 & 10.4  & 15.3 & 1.6e8 & 220  \\
NGC4649 & 1 & 2.42 & 17.17 & 2.00 & 1.30 & 0.15 & 3.74 & 16.2  & 15.3 & 1.2e9 & 370  \\
NGC4697 & 1 & 2.12 & 16.93 &24.86 & 1.04 & 0.74 & 3.58 &  6.8 & 10.5 & 1.4e8 & 210  \\
NGC4874 & 1 & 3.08 & 19.18 & 2.33 & 1.37 & 0.13 & 4.44 & 15.0  & 93.3 & 4.3e8 & 280  \\
NGC4889 & 1 & 2.88 & 18.01 & 2.61 & 1.35 & 0.05 & 4.15 & 11.2  & 93.3 & 8.7e8 & 340  \\
NGC5813 & 1 & 2.04 & 16.42 & 2.15 & 1.33 & 0.08 & 3.82 &  7.1 & 28.3 & 1.2e8 & 200  \\
NGC6166 & 1 & 3.08 & 19.35 & 3.32 & 0.99 & 0.08 & 4.49 & 15.6  &112.5 & 1.1e9 &
360  \\
NGC221  & 1 &-0.26 & 11.77 & 0.98 & 1.36 & 0.01 & 2.18 &  2.3 &  0.8 & \multicolumn{2}{c}{$DF<0$} \\
NGC1316 & 3 & 1.55 & 14.43 & 1.16 & 1.00 & 0.00 & 3.84 &  2.6 & 17.9 & \multicolumn{2}{c}{$DF<0$} \\
NGC1400 & 1 & 1.54 & 15.41 & 1.39 & 1.32 & 0.00 & 3.60 & 10.7  & 21.5 & \multicolumn{2}{c}{$DF<0$} \\
NGC1700 & 1 & 1.19 & 13.95 & 0.90 & 1.30 & 0.00 & 3.61 &  4.0 & 35.5 & \multicolumn{2}{c}{$DF<0$} \\
NGC2636 & 1 & 1.17 & 15.68 & 1.84 & 1.14 & 0.04 & 2.86 &  3.0 & 33.5 & \multicolumn{2}{c}{$DF<0$} \\
NGC3608 & 1 & 1.44 & 15.45 & 1.05 & 1.33 & 0.00 & 3.54 &  7.0 & 20.3 & \multicolumn{2}{c}{$DF<0$} \\
NGC4472 & 1 & 2.25 & 16.66 & 2.08 & 1.17 & 0.04 & 3.89 &  9.2 & 15.3 & \multicolumn{2}{c}{$DF<0$} \\
NGC4552 & 1 & 1.68 & 15.41 & 1.48 & 1.30 & 0.00 & 3.35 &  7.7 & 15.3 & \multicolumn{2}{c}{$DF<0$} \\
NGC7768 & 1 & 2.30 & 16.99 & 1.92 & 1.21 & 0.00 & 4.18 &  9.5 &103.1 & \multicolumn{2}{c}{$DF<0$} \\
NGC221V & 1 & 2.43 & 20.42 & 1.72 & 3.55 & 1.21 & 2.18 &  1.3 & 19.2 & \multicolumn{2}{c}{$\sigma\e$ is affected by $M_\bullet$} \\
NGC4742 & 2 & 1.93 & 16.69 &48.60 & 1.99 & 1.09 & 2.85 &  1.8 & 12.5 & \multicolumn{2}{c}{$\sigma\e$ is affected by $M_\bullet$} \\
\hline
\end{tabular}
\medskip
\caption{Galaxy Sample.
The second column gives the Hubble types of the galaxies:
1=E, 2=E/S0 or S0, 3=Sa or Sb, 4=dE or dSph.
The $r\b$ and $\mu\b$ are the break radius and break surface
brightness (corrected for Galactic extinction, in $V$ mag ${\rm arcsec}^{-2}$) 
in the Nuker law (equation \ref{eq:nukerlaw}); and $\alpha$, $\beta$ and $\gamma$ are the exponents
describing the sharpness of the break, the outer and inner slopes of the
Nuker law.
The $r_{\rm e}$ is the half-light radius and $\Upsilon_V$ is the mass-to-light
ratio in the $V$ band.
$M_\bullet$ is the central black hole mass and $\sigma\e$ is the
luminosity-weighted line-of-sight velocity dispersion inside $r_{\rm e}$. 
All the parameters except $M_\bullet$ and $\sigma\e$ are adapted from
Faber et al. (1997) where $H_0=80\kms~{\rm Mpc}^{-1}$.
The $\sigma\e$ comes from the calculation based on a spherical and
isotropic model in this paper and $M_\bullet$ is obtained from the correlation
of BH mass with galactic velocity dispersion (equation \ref{eq:msigma}).
The 9 galaxies with ``$DF<0$'' are deleted from our study because their
distribution functions obtained from the Eddington formula are negative  
(equation \ref{eq:Eddington}); and so are the last two galaxies with $\gamma>1$
because their $\sigma\e$ are significantly affected by the gravitational
potential of central BHs (see \S~\ref{sec:galpro}).}
\label{tab:tab1}
\end{minipage}
\end{table*}

\end{document}